\documentclass[aps,pre,reprint,amsmath,amssymb,superscriptaddress,showpacs,floatfix]{revtex4-1}
\usepackage{graphicx}
\usepackage{dcolumn}
\usepackage{ulem}
\usepackage{bm}
\usepackage{booktabs}
\usepackage[para]{threeparttable}
\usepackage{tabularx}
\usepackage[colorlinks, allcolors=blue]{hyperref}
\usepackage[usenames, dvipsnames]{color}

\begin{document}

\title{Spin Peierls transition of $J_{1}-J_{2}$ and extended models with ferromagnetic $J_{1}$.\\ Sublattice dimerization and thermodynamics of zigzag chains in $\beta$-TeVO$_{4}$}

\author{Manodip Routh}
\affiliation{S. N. Bose National Centre for Basic Sciences, Block - JD, Sector - III, Salt Lake, Kolkata - 700106, India}

\author{Sudip Kumar Saha}
\affiliation{S. N. Bose National Centre for Basic Sciences, Block - JD, Sector - III, Salt Lake, Kolkata - 700106, India}

\author{Manoranjan Kumar}
\email{manoranjan.kumar@bose.res.in}
\affiliation{S. N. Bose National Centre for Basic Sciences, Block - JD, Sector - III, Salt Lake, Kolkata - 700106, India}

\author{Zolt\'an G. Soos}
\email{soos@princeton.edu}
\affiliation{Department of Chemistry, Princeton University, Princeton, New Jersey 08544, USA}

\date{\today}

\begin{abstract}
The spin$-1/2$ chain with ferromagnetic exchange $J_1 < 0$ between first neighbors and antiferromagnetic $J_2 > 0$ between second neighbors supports two spin-Peierls (SP) instabilities depending on the frustration $\alpha = J_2/\vert J_1\vert$. Instead of chain dimerization with two spins per unit cell, $J_1-J_2$ models with $\alpha > 0.65$ and linear spin-phonon coupling are unconditionally unstable to sublattice dimerization with four spins per unit cell. Unequal $J_1$ to neighbors to the right and left extends the model to gapped ($\gamma > 0$) chains with conditional SP transitions at $T_{SP}$ to dimerized sublattices and a weaker specific heat $C(T)$ anomaly. The spin susceptibility $\chi(T)$ and $C(T)$ are obtained in the thermodynamic limit by a combination of exact diagonalization of small systems with $\alpha > 0.65$ and density matrix renormalization group (DMRG) calculations of systems up to $N \sim 100$ spins. Both $J_1-J_2$ and $\gamma > 0$ models account quantitatively for $\chi(T)$ and $C(T)$ in the paramagnetic phase of $\beta$-TeVO$_{4}$ for $T > 8$ K, but lower $T$ indicates a gapped chain instead of a $J_1-J_2$ model as previously thought. The same parameters and $T_{SP} = 4.6$ K generate a $C(T)/T$ anomaly that reproduces the anomaly at the $4.6$ K transition of $\beta$-TeVO$_{4}$, but not the weak $\chi(T)$ signature.
\end{abstract}

\maketitle

\section{\label{sec1} Introduction}
The $J_{1}-J_{2}$ model [Eq.~(\ref{Eq1}) with $\delta = 0$] is a spin$-1/2$ chain with isotropic exchange $J_1$ and $J_2$ between first and second neighbors, respectively, and one spin per unit cell. The 1D model with antiferromagnetic $J_2 > 0$ is frustrated for either sign of $J_1$, quantified by the parameter $ \alpha = J_{2}/\vert J_{1} \vert > 0 $. Theoretical interest initially focused ~\cite{Sandvik2010,lecheminant2004one,okamoto1992fluid,
majumdar1969next,affleck1989critical,chitra1995density,eggert1996numerical,
sirker2010thermodynamics,soos2016numerical,*kumar2013decoupled} on quantum ($T = 0$) phases in the $J_{1} > 0$ sector and the critical point~\cite{okamoto1992fluid} $\alpha_{c} = 0.2411$ between the gapless linear Heisenberg antiferromagnet (HAF) at $\alpha = 0$ and the gapped Majumdar-Ghosh point~\cite{majumdar1969next} at $\alpha = 1/2$. Field theoretical and numerical methods were developed for the transition. The $J_{1}-J_{2}$ model supports a host of exotic quantum phases (vector chiral, multipolar, nematic,... etc.) in an applied magnetic field or when also considering anisotropic or antisymmetric exchange.~\cite{chubukov1991chiral,heidrich2006frustrated, dmitriev2008weakly,hikihara2008vector,sirker2011j,furukawa2012ground,
*furukawa2010unconventional} However, exact thermodynamics~\cite{johnston2000thermodynamics} is limited to $J_2 = 0$. 

As pointed out by Hase et al.~\cite{hase2004} the thermodynamics of quasi-1D materials is directly and sometimes semi-quantitatively related to $J_{1}-J_{2}$ models with either sign of $J_{1}$. The topics of the present study are ferromagnetic exchange and thermodynamics. Among others, spin-$1/2$ Cu(II) chains with $J_1 < 0$ have recently been identified among cupric oxides with an estimated $\alpha$ ranging from  ~\cite{dutton2012dominant, *kumar2013spin} $\alpha \approx 0$ in Ba$_{3}$Cu$_{3}$In$_{4}$O$_{12}$ or Ba$_{3}$Cu$_{3}$Sc$_{4}$O$_{12}$ to $\alpha \approx 0.5$ in (N$_{2}$H$_{5}$)CuCl$_{3}$,~\cite{maeshima2003magnetic} LiCu$_{2}$O$_{2}$,~\cite{park2007ferroelectricity} LiCuSbO$_{4}$,~\cite{dutton2012q} LiCuVO$_{4}$,~\cite{mourigal2012evidence} and Rb$_{2}$Cu$_{2}$Mo$_{3}$O$_{12}$.~\cite{yagi2017nmr,hase2004} The unpaired spin is in the $Cu^{2+}$ 3d$_{x2-y2}$ orbital with lobes pointing to the bridging O$^{2-}$ ions. Interchain and other interactions inevitably become important at low $T$ and almost always drive phase transitions that are difficult to model.

\begin{figure}[hbtp]
\begin{center}
\includegraphics[width=\columnwidth]{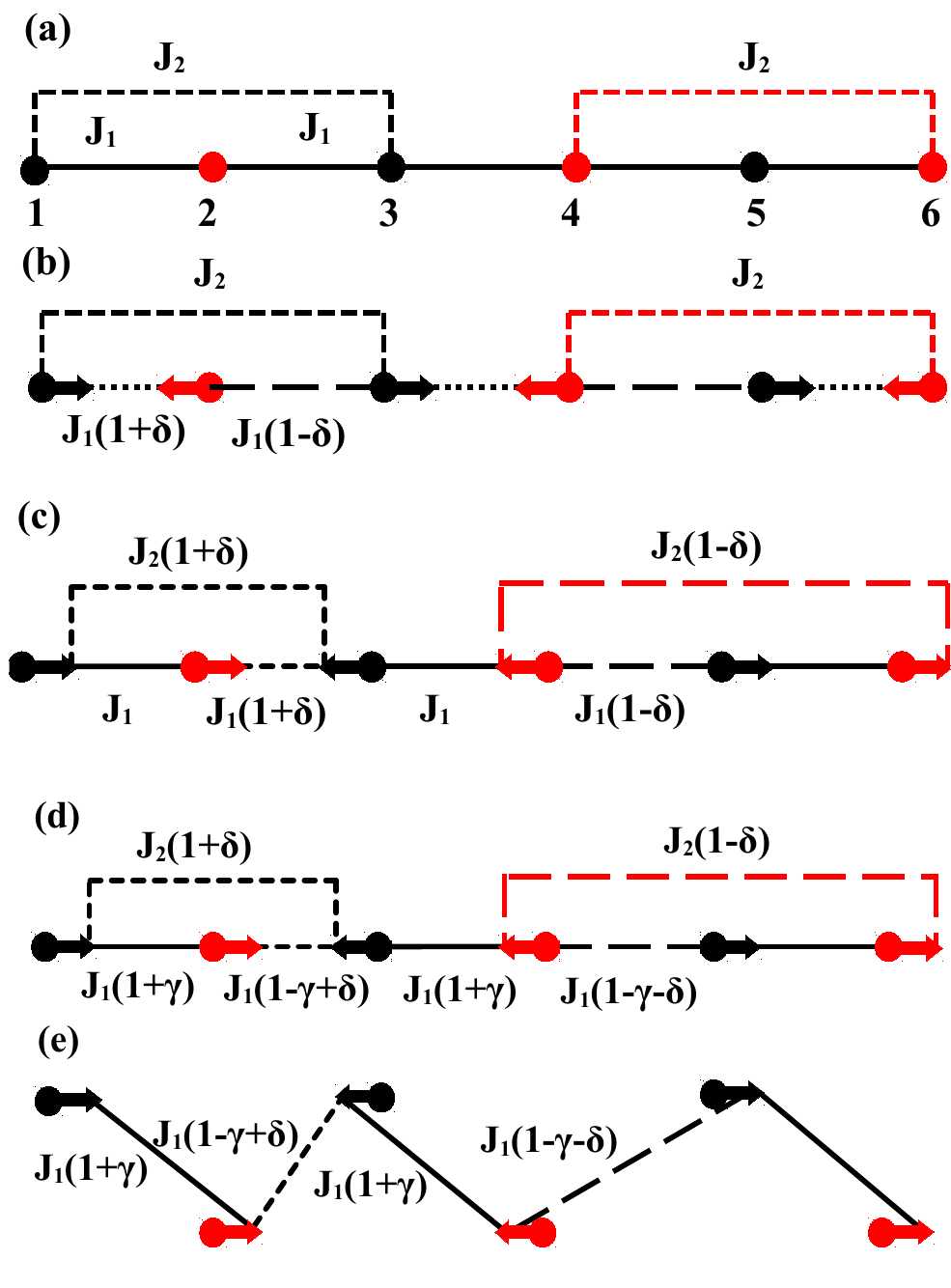}
\caption{(a)-(c): Schematic representations of the $J_1-J_2$ model, Eq.~(\ref{Eq1}). Exchanges change by $(1 + \delta)$ or $(1 - \delta)$ for arrows that decrease or increase the bond length. (a) Regular chain with equally spaced spins; (b) dimerized chain with alternating $J_1(1 \pm \delta)$ and unchanged $J_2$; (c) sublattice dimerized chain with alternating $J_2(1 \pm \delta)$.(d)(e) Extended $J_1-J_2$ model with collinear (d) and zigzag (e) sublattice dimerization. Parallel arrow do not change $J_1(1+\gamma)$ at every other bond; the other bond alternates as $J_1(1-\gamma \pm \delta)$. The extended zigzag model at $\delta = 0$ is a ladder with rails $J_2$ and rungs $J_1(1 \pm \gamma)$.}
\label{fig1}
\end{center}
\end{figure}

The $J_{1}-J_{2}$ model is defined on a regular chain of equally spaced sites as sketched in Fig.~\ref{fig1}(a). It is subject to a spin-Peierls (SP) instability on a deformable chain with spin-phonon coupling. The dimerized chain at $T < T_{SP}$ in Fig.~\ref{fig1}(b) has alternating $J_{1}(1 \pm \delta)$ along the chain and increasing $\delta(T)$ on cooling to $T = 0$. At the mean field level, the spin gap $\Delta$($\delta(T)$) of the HAF at $T < T_{SP}$ follows~\cite{jacobs1976spin} the BCS gap equation for superconductors.

We discuss in this paper the SP transition of the ferromagnetic $J_{1}-J_{2}$ model, which differs in several ways from the antiferromagnetic case and has not been reported previously. The $J_{1} < 0$ model has two SP transitions depending on $\alpha$. Intermediate $\alpha \sim 0.5$ again leads to chain dimerization at $T < T_{SP}$. At larger $\alpha \sim 1$, the transition is instead to sublattice dimerization in 
Fig.~\ref{fig1}(c) with alternating $J_{2}(1 \pm \delta)$, four spins per unit cell and four equivalent ground states. Sublattice dimerization is readily understood at $\alpha > 1$ when the $J_{1}-J_{2}$ model corresponds to weakly coupled HAFs on sublattices of odd and even numbered sites. The instability to sublattice dimerization is presented in Sec.~\ref{sec2}.

The spin Hamiltonian of the ferromagnetic $J_{1}-J_{2}$ model with periodic boundary conditions, $\vert J_{1} \vert = 1$ as the energy unit, $J_{2} = \alpha$ and \textit{chain} dimerization $ \delta $ in Fig.~\ref{fig1}(b) is
\begin{equation}
H(\alpha, \delta) = \sum_{r}-(1-\delta(-1)^{r}) \vec{S}_{r}\cdot\vec{S}_{r+1} + \alpha \vec{S}_{r}\cdot\vec{S}_{r+2}
\label{Eq1}
\end{equation}
$H(\alpha,0)$ is the standard $J_{1}-J_{2}$ model with equally spaced spins, one spin per unit cell and inversion symmetry at sites. The ground state of $H(\alpha,0)$ is ferromagnetic up to the exact quantum critical point,~\cite{hamada1988exact} $\alpha_{c} = 1/4$, and is a singlet, $S = 0$, for $\alpha \geqslant \alpha_{c}$, the sector we consider in this paper. Field theory distinguishes sharply between gapless quantum phases of the $J_{1}-J_{2}$ model with nondegenerate ground states and gapped phases with doubly degenerate ground states.~\cite{allen1997non} The incommensurate (IC) phase at $ \alpha \geqslant \alpha_{c} $ is gapped. The singlet-triplet gap $\Delta(\alpha) $ is exponentially small,~\cite{itoi2001strongly} however, recently estimated~\cite{Agrapidis} as $\Delta(\alpha)  < 10^{-3}$. The decoupled phase \cite{soos2016numerical,*kumar2013decoupled} at $\alpha > 0.80$ is gapless and reduces to HAFs on sublattices in the limit $J_1 = 0$. The HAF is by far the best characterized spin-$1/2$ chian; it corresponds to $\alpha = \delta = 0$ in Eq.~(\ref{Eq1}) with antiferromagnetic $J_1 > 0$.

The $J_{1}-J_{2}$ model has been applied to quasi-1D materials, as mentioned above, whose crystal structure indicates chains with two spins per unit cell and no inversion symmetry at sites before any consideration of spin-phonon coupling. The extended $J_{1}-J_{2}$ model in Fig.~\ref{fig1}(d) allows for the possibility of different exchange with right and left neighbors, which we take as $J_{1}(1 \pm \gamma)$. The chain $H(\alpha,\gamma)$ with $\gamma$ instead of $\delta$ in Eq.~(\ref{Eq1}) has lower symmetry and no inversion at sites. The SP transition of the extended model is to sublattice dimerization with four spins per unit cell. Sublattice dimerization in Figs.~\ref{fig1}(c) and ~\ref{fig1}(d) holds for both the $J_{1}-J_{2}$ and extended models at $\alpha \sim 1$. The thermodynamics of the extended model at $T > T_{SP}$ are given by Eq.~(\ref{Eq1}) with $\delta = \gamma$.

The $J_{1}-J_{2}$ and extended models are strongly correlated spin systems with frustration $\alpha$. Their $T = 0$ properties have been investigated by multiple mathematical methods. Fewer methods are available for thermodynamics, especially at low $T$. Quantum Monte Carlo is not applicable to frustrated systems. We obtain~\cite{saha2019hybrid} the thermodynamic limit in Sec.~\ref{sec3} by exact diagonalization (ED) up to system size $N$ followed by density matrix renormalization group (DMRG) calculations at progressively larger $N$. The antiferromagnetic $J_1-J_2$ model has been studied using the transfer matrix renormalization group (TMRG)~\cite{Maeshima2000} and $T$ dependent DMRG.~\cite{Feiguin2005} The magnetic susceptibility $\chi(T)$ and specific heat $C(T)$ are shown~\cite{Maeshima2000,Feiguin2005} down to $T/J_{1} = 0.05$ and differences appear below $0.10$. We recently showed~\cite{saha2021} that ED/DMRG is accurate down to $T/J_{1} \sim 0.02$ for the $J_{1}-J_{2}$ model with $J_{1} > 0$. All the methods are open to improvements.

The most extensively characterized SP systems, both with $J_1 > 0$, are the inorganic crystal~\cite{Hase1993PRL,*Hase1993PRB} CuGeO$_3$ with $\alpha = 0.35$ that was studied intensively in the 90s and the organic molecular crystal~\cite{jacobs1976spin} TTF-CuS$_4$C$_4$(CF$_3$)$_4$ with $\alpha = 0$ (HAF) modeled a decade earlier. Quantitative $\chi(T)$ modeling was achieved~\cite{Fabricius1998, jacobs1976spin} for both for $T > T_{SP}$ using correlated spin states. Large CuGeO$_3$ crystals suitable for inelastic neutron scattering~\cite{Nishi_1994, Lussier_1996, Martin_1996, Regnault_1996} revealed the limitations of mean field theory for $T < T_{SP}$. The $T$ dependence of the spin gap did not follow the BCS gap equation, a puzzle that is unresolved in the 2002 review~\cite{Uchinokura_2002} of Uchinokura. The SP transition and $\chi(T)$ have been modeled by TMRG~\cite{Raupach_1999} and by ED/DMRG~\cite{saha_2020} , and are compared in the Discussion section below. Correlated states~\cite{saha_2020} account for the neutron data and fit the specific heat anomaly of CuGeO$_{3}$ as well as $\chi(T)$. They also return an internally consistent spin gap for the HAF, whose $T$ dependence differs from BCS.

The magnetic properties of $\beta$-TeVO$_4$ have been actively studied for a decade.~\cite{savina2011magnetic, pregelj2015, weickert2016magnetic,savina2015features} it is a novel quasi-1D material in several ways. The crystal structure~\cite{savina2011magnetic} indicates zigzag spin$-1/2$ chains, as sketched in Fig.~\ref{fig1}(e), along the $c$ axis in the $bc$ plane. Three transitions below $10$ K have been discussed~\cite{savina2011magnetic, pregelj2015, weickert2016magnetic} in terms of the $T = 0$ phases of the ferromagnetic $J_1-J_2$ model, which has generally been invoked for the paramagnetic phase. The unpaired electron is in the $V^{4+}$ $3d_{xy}$ orbital~\cite{saul2014density,kivelson1964esr}  with lobes pointing at approximately $\pm \pi/4$ from the bridging $O^{2-}$. The molar spin susceptibility $\chi(T)$ and spin specific heat $C(T)$ have been fit~\cite{ savina2011magnetic,pregelj2015,weickert2016magnetic,savina2015heat} with slightly different $J_{1}$ and $\alpha \sim 0.77$ to $1$. The $4.6$ K transition to an antiferromagnetic phase has not been modeled. All three transitions have been 
qualitatively
attributed to exchange interactions, both ferro and antiferromagnetic, between spins in adjacent chains.~\cite{savina2011magnetic,pregelj2015,weickert2016magnetic} We find below that the ferromagnetic $J_1-J_2$ model does \textit{not} account for magnetic data at low $T$, which point instead to a gapped system.

The extended model, Eq.~(\ref{Eq1}) with $\gamma$ instead of $\delta$, has two spins per unit cell and a finite singlet-triplet gap $\Delta(\alpha, \gamma) > 0$ that increases with $\gamma$. The singlet ground state is nondegenerate. The extended $J_1-J_2$ model provides a convenient and controlled way to study gapped chains. We obtain the thermodynamics of the 1D model, $H(\alpha,\gamma)$, quantitatively for $T > 0.03\vert J_1 \vert$. The parameters $ \alpha $ and $ \gamma $ are chosen with $\beta$-TeVO$_4$ in mind. The joint analysis of thermodynamic quantities shows that $H(\alpha,\gamma)$ with $\gamma \sim 0.15$ is a first approximation to $\beta$-TeVO$_4$.

It is essential to distinguish between 1D models such as Eq.~(\ref{Eq1}) for rigid chains and quasi-1D materials. Models with isotropic exchange and equally spaced spins are of great mathematical interest. Isotropic exchange is the dominant interaction that governs the thermodynamics. However, Eq.~(\ref{Eq1}) is \textit{approximate and incomplete} quite aside from neglecting all interchain interactions and spin-phonon coupling. Approximate because spin-orbit coupling generates corrections to isotropic exchange and leads to g-tensors in an applied magnetic field. Incomplete because dipolar interactions between electronic spins are neglected as well as hyperfine interactions with nuclear spins. Small interactions, both intra and interchain, become important at low $T$. The distinction between models and materials is particularly sharp for $\beta$-TeVO$_4$. The $J_1-J_2$ model has one spin per unit cell and inversion symmetry at sites. The $\beta$-TeVO$_4$ symmetry is far lower.~\cite{savina2011magnetic} There are two zigzag chains, each with two spins per unit cell. The inversion centers are between chains and interchange the chains.

The paper is organized as follows. We discuss in Sec.~\ref{sec2} the instability of the $J_1-J_2$ model in the singlet sector to chain or sublattice dimerization with increasing $\alpha$. The instability of the extended ($\gamma > 0$) model to sublattice dimerization is conditional. The thermodynamics of $H(\alpha,\gamma)$ is obtained in Sec.~\ref{sec3}. The spin susceptibility $ \chi(T,\alpha,\gamma) $ and specific heat $C(T,\alpha,\gamma)$ depend strongly on $ \gamma $ at low $  T$. The SP transition to dimerized sublattices is modeled conventionally~\cite{su1980soliton} in Sec.~\ref{sec4}, again contrasting $J_1-J_2$ and extended models, in terms of linear spin-phonon coupling and a harmonic lattice. We show in Sec.~\ref{sec5} that the extended $J_1-J_2$ model accounts for $\beta$-TeVO$_4$ thermodynamics in the paramagnetic phase and model published $\chi(T)$ and $C(T)$ data down to $2$ K, well below the $4.6$ K transition. The transition combines a clear $C(T)$ anomaly with a very weak $\chi(T)$ signature. Section~\ref{sec6} provides a brief discussion.


\section{\label{sec2} Electronic Instability}
Peierls pointed out the ground-state instability of a 1D metal. At constant bandwidth $4t$, linear electron-phonon coupling opens a gap at the Fermi wave vector $k_{F}$ in lattices with a harmonic potential. Dimerization $t(1 \pm \delta)$ lowers the ground-state energy of the half-filled band by opening a gap at $k_F = \pm \pi/2$. The electronic instability applies to correlated systems or to linear spin-phonon coupling in the HAF or other spin chains. Any small modulation of exchanges that conserves total $J_1$ and total $J_2$ is admissible. Periodic boundary conditions correspond to constant length or 1D volume per spin. The instabilities of the antiferromagnetic $J_1-J_2$ model to dimerization at frustration $\alpha$ have been discussed in detail.

Our motivation here is different. The instability of the ferromagnetic $J_1-J_2$ model in the singlet sector with $\alpha > \alpha_{c} = 1/4 $ is to sublattice dimerization at large $  \alpha$ when exchange between second neighbors dominates and to chain dimerization at intermediate $  \alpha$. The electronic stabilization follows from $H(\alpha,\delta)$, Eq.~(\ref{Eq1}), with modulated exchanges in Fig.~\ref{fig1}(b) for chain dimerization and Fig.~\ref{fig1}(c) for sublattice dimerization. We begin with the instabilities of the $J_1-J_2$ model. When sublattice dimerization leads to greater stabilization, we compare the instabilities of the $J_1-J_2$ and extended ($\gamma > 0$) models.

Linear spin-phonon coupling increases the exchange in shortened bonds by $(1 + \delta)$ and decreases it by $(1 - \delta)$ in lengthened bonds. Total exchange is conserved separately for first and second neighbors. We obtain the ground state of $H(\alpha,\delta)$ in the thermodynamic limit by ED and DMRG solution of finite chains with periodic boundary conditions. The electronic stabilization per spin of the $J_1-J_2$ model on 
chain dimerization is
\begin{equation}
\Delta E^{chain}(\alpha,\delta) = E_{0}(\alpha,\delta) - E_{0}(\alpha,0)
\label{Eq2}
\end{equation}
where $E_0$ is the ground state energy per spin.

Sublattice dimerization in Fig.~\ref{fig1}(c) leads to alternating $J_2 = \alpha(1 \pm \delta)$ and to $J_{1} = -1$ for every left or right neighbor. Exchange to the other neighbor alternates as $-(1 \pm \delta)$. The Hamiltonian of the sublattice dimerized $J_1-J_2$ model ($\gamma = 0$) is
\begin{eqnarray}
H_{S}(\alpha,\gamma, \delta)&=& \sum_{r} -\left( 1-\gamma(-1)^{r} -\delta cos\frac{\pi r}{2}\right) \vec{S}_{r}\cdot \vec{S}_{r+1}\nonumber
\\&+& \alpha \left(1+\sqrt{2}\delta sin\frac{\pi(2r-1)}{4}\right) \vec{S}_{r}\cdot\vec{S}_{r+2} 
\label{Eq3} 
\end{eqnarray}
Sublattice dimerization of the extended model has $\gamma > 0$. We solve $H_{S}(\alpha,0,\delta)$ in finite chains with periodic boundary conditions to obtain the electronic stabilization of the $J_{1}-J_{2}$ model for sublattice dimerization
\begin{equation}
\Delta E^{sub}(\alpha,0,\delta) = E_{S}(\alpha,0,\delta) - E_{S}(\alpha,0,0)
\label{Eq4}
\end{equation}
where $E_{S}$ is the ground state energy per spin.

Every spin is displaced by $\pm \delta $ for either chain or sublattice dimerization. The elastic energy is equal at the level of Einstein phonons, but not in more realistic treatments. In a 1D approximation, the frequency of the optical phonon for chain dimerization is higher, hence stiffer, than that of the $q = \pi/4$ phonon for sublattice dimerization. Electronic stabilization at equal $ \delta $ is a rough estimate of the preferred deformation from equal spacing. 

\begin{figure}[hbtp]
\begin{center}
\includegraphics[width=\columnwidth]{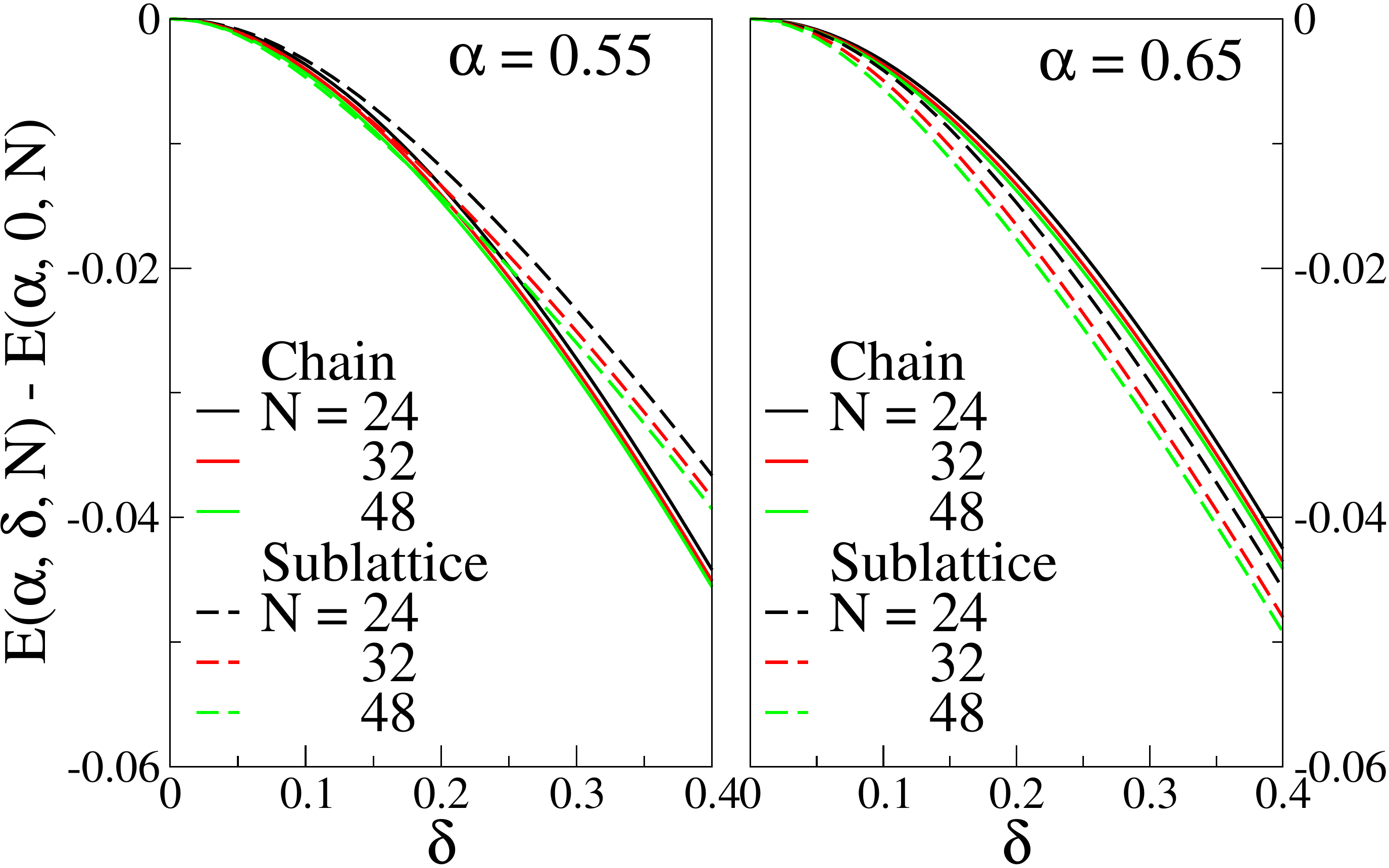}
\caption{Ground state stabilization per site of the ferromagnetic $J_1-J_2$ model with
$N$ spins and $\alpha = 0.55$ or $0.65$. Solid lines are Eq.~(\ref{Eq2}) for chain dimerization $­-(1\pm \delta)$ and $\alpha$ for second neighbors. Dashed lines are Eq.~(\ref{Eq4}) with $\gamma = 0$ for sublattice dimerization with $ \alpha(1\pm \delta)$ for second neighbors, $-1$ for every other first neighbor and $-(1 \pm \delta)$ for the rest.}
\label{fig2}
\end{center}
\end{figure}

We compare electronic stabilization, Eqs.~(\ref{Eq2}) and~(\ref{Eq4}), at $\alpha = 0.55$ and $0.65$ in the left and right panels of Fig.~\ref{fig2} using ED for $N = 24$ and DMRG for $N = 32$ and $48$. Sublattice dimerization lowers the energy more for $ \alpha \geqslant 0.65$, well below $J_{2} = -J_{1}$. More precisely, increasing $\alpha$ from $0.55$ to $0.65$ \textit{increases} $\Delta E^{chain}$ by $8\%$ at $\delta = 0.1$ and $N = 48$ while it decreases $\Delta E^{sub}$ by $25\%$. The preference for sublattice dimerization increases rapidly for larger $ \alpha $.

Previous SP systems with $J_{1} > 0$ were unconditionally unstable to chain dimerization. The ferromagnetic $J_1-J_2$ model supports chain dimerization between $\alpha_{c} = 1/4$ and $0.55$ and sublattice dimerization for $\alpha > 0.65$. We have not sought the precise boundary, which also depends on the lattice. Since sublattice dimerization leads to four spins per unit cell, the instability also applies to the extended model $H(\alpha,\gamma)$ with $\gamma > 0$ instead of $\delta$ in Eq.~(\ref{Eq1}) and two spins per unit cell. Without loss of generality, we take $J_{1} = - (1 \pm \gamma)$ for the left and right neighbors, respectively. Sublattice dimerization does not change $J_1$ with one neighbor and modulates it with the other. The ground state energy of Eq.~(\ref{Eq3}) in finite chains is lower for unchanged  $J_1 = - (1 + \gamma)$ and alternating $J_1 = -(1 -\gamma \pm \delta)$ than for unchanged $J_{1} = -(1 - \gamma)$ and alternating $J_1 = -(1 + \gamma \pm \delta)$. 

The zigzag chain or two-rail ladder in Fig.~\ref{fig1}(e) has rails with exchange $J_2$ and rungs $J_1$. The arrows indicate sublattice dimerization leading to alternating $\alpha(1 \pm \delta)$ along the rails. In the extended model, parallel arrows at adjacent sites still return $J_1 = - (1 + \gamma)$  at every other rung. The $J_1$ modulation of the remaining rungs differs from $J_1 = -(1 - \gamma \pm \delta)$ by geometric factors since antiparallel arrows are not along rungs. This level of detail is premature in our opinion since exchange interactions depend on orbital pathways as well as on distance. 

All calculations from here on are performed on chains with collinear sublattice dimerization. Equation~(\ref{Eq3}) with $\gamma = 0$ is the $J_1-J_2$ model in Fig. ~\ref{fig1}(c) with unchanged $J_1 = -1$ at every other bond and modulated $J_{1} = -(1 \pm \delta)$ at the other bond. The extended model in Fig.~\ref{fig1}(d) has $\gamma > 0$ in Eq.~(\ref{Eq3}). The unchanged $J_1$ is  $-(1+\gamma)$; the modulated $J_1$ is $ -(1-\gamma \pm \delta)$.

As already mentioned, the singlet-triplet gap $\Delta(\alpha,\gamma)$ of the extended model increases with $\gamma $ while the gap $\Delta(\alpha)$ of the ferromagnetic $J_1-J_2$ model is zero in the decoupled phase and exponentially small in the IC phase. Finite $\Delta(\alpha,\gamma)$ leads to a conditional SP transition to dimerized sublattices. The SP instability to sublattice dimerization is driven by the electronic force constant, the curvature of the ground state energy per site
\begin{equation}
E_{0}^{\prime \prime}(\alpha,\gamma)=\left(\frac{\partial^{2}E_{0}(\alpha,\gamma,\delta)}{\partial\delta^{2}}\right)_0.
\label{Eq5}
\end{equation}

$E_{0}(\alpha,\gamma,\delta)$ is the per site energy of Eq.~(\ref{Eq3}), the $J_1-J_2$ model when $\gamma = 0$ and the extended model when $\gamma > 0$. The curvature is negative since $E_{0}(\alpha,\gamma,\delta)$ has a maximum at $\delta = 0$. The instability is conditional if $E_{0}^{\prime \prime}(\alpha,\gamma,N)$ is finite in the thermodynamic limit. The size dependence of the curvature is listed in Table~\ref{tab1} up to $N = 64$ for $ \alpha = 0.75$ and $1$, $\gamma = 0$ and $0.15$. Numerical derivatives are small differences that require accurate $E_{0}(\alpha,\gamma,\delta,N)$. The $\gamma > 0$ entries converge with system size while the $\gamma = 0$ entries increase steadily with $N$ as expected. Much larger $N$ would be required to probe $\Delta(\alpha)$ in the IC phase. To a first approximation, the curvatures in Table~\ref{tab1} are proportional to $ \alpha $, the antiferromagnet exchange between neighbors within a sublattice.

\begin{table}[hbtp]
\caption{\label{tab1}
Size dependence of the curvature $ E^{\prime \prime}(\alpha,\gamma) $ in Eq.~(\ref{Eq5}) of chains with $\alpha_{1}= 0.75$, $\alpha_{2} = 1.0$, $N$ spins, and $\gamma_{1} = 0$, $\gamma_{2} = 0.15$ in Eq.~(\ref{Eq1}) with $\delta = \gamma$.}
\begin{ruledtabular}
\begin{tabular}{ c c c c c }
$N$ & $E^{\prime \prime}(\alpha_{1},\gamma_{1}) $ & $E^{\prime \prime}(\alpha_{1},\gamma_{2}) $ & $E^{\prime \prime}(\alpha_{2},\gamma_{1}) $ & $E^{\prime \prime}(\alpha_{2},\gamma_{2}) $ \\
\hline
24 & $-$1.511 & $-$1.324 & $-$2.103 & $-$1.968\\
32 & $-$1.976 & $-$1.498 & $-$2.704 & $-$2.278\\
48 & $-$2.940 & $-$1.501 & $-$3.877 & $-$2.298\\
64 & $-$4.085 & $-$1.504 & $-$5.126 & $-$2.304\\
\end{tabular}
\end{ruledtabular}
\end{table}
To summarize Sec.~\ref{sec2}, sublattice dimerization is the SP instability of the ferromagnetic $J_1-J_2$ model with $\alpha > 0.65$. The spin Hamiltonian is $H_{S}(\alpha, \gamma, \delta)$, Eq.~(\ref{Eq3}), with $\gamma = 0$ in the $J_I-J_2$ model and $\gamma > 0$ in the extended model. We have $\delta =0$ at $T > T_{SP}$ in the paramagnetic phase and $\delta > 0$ on sublattice dimerization. The instability is unconditional in the $J_1-J_2$ model and conditional in the extended model. We obtain the low $T$ thermodynamics of $H_{S}(\alpha,\gamma,0)$ in Sec.~\ref{sec3} and the SP transition of both models to sublattice dimerization in Sec.~\ref{sec4}.
 
\section{\label{sec3} Thermodynamics}
Spin chains with isotropic exchange have $2^{N}$ spin states at system size $N$. Both the total spin $S \leqslant N/2$ and its Zeeman component $S^Z$ are conserved. We obtain the low $T$ thermodynamics of $H_S(\alpha,\gamma,0)$ for parameters that are relevant to the paramagnetic phase of $\beta$-TeVO$_{4}$.

ED yields the full spectrum ${E(\alpha,\gamma,N)}$ up to $N$ = 24 at $\gamma = 0$ and to $N$ = 20 for $\gamma > 0$ and lower translational symmetry. DMRG calculations for larger $N$ and periodic boundary conditions are performed in sectors with fixed $S^Z$ as discussed in Ref.~\cite{saha2019hybrid}. The superblock has four new sites in addition to the left and right blocks. The dimension of the superblock Hamiltonian is $m^{2}4^{2}$ where $m$ (usually 400) is the number of eigenvectors that correspond to the highest eigenvalues of the system block density matrix. The $S^Z = 0$ ground state $E_1(N)$ is taken as zero energy; states $j > 1$ have excitation energy $E_j(N) > 0$. To obtain the lowest excitations very accurately, we first target the lowest $l = 5-10$ states of the superblock. The second calculation has $l > 100$ (usually 200). The entire spectrum is red shifted by an approximately constant amount because the density matrix has projections from many excited states. Accordingly, we shift the spectrum back by a constant and use the first calculation for the lowest excitations.

We then introduce an energy cutoff $W(N)$ and construct the canonical partition function $Q_C(T,N)$ using all $R(N)$ states with $E_j(N) \leqslant W(N)$. The entropy $S_C(T,N)$, magnetic susceptibility $\chi_{C}(T,N)$ and other quantities are obtained using $Q_C(T,N)$ with $R(N)$ states. The thermodynamic limit at system size $N$ is reached by increasing $W(N)$ until the maximum of $S_C(T,N)/T$ at $T(N)$ has converged or almost converged, typically for $R(N) \sim 10^3$ states. Since we have $S^{\prime}(T_m) = S(T_m)/T_m$ at the maximum of $S(T)/T$, size convergence at the maximum returns the thermodynamic limit of $S^{\prime}(T)$ at $T = T(N)$. Finite size effects are suppressed in a narrow range around $T(N)$ before truncation takes its toll.

Table~\ref{tab2} lists $T(N)$ for the indicated $\alpha$, $\gamma$ in Eq.~(\ref{Eq3}) with $\delta = 0$. Converged thermodynamics are accessible for $T > 0.03$ at systems sizes up to $N = 96$. The points $T(N)$ are closely spaced, and additional points can be computed if desired. Larger $N \sim 200$ is accessible with greater effort. The size limit is set by the density of correlated states of the model. Extrapolation to lower $T$ is possible, but not in general to $T = 0$ where, for example, the HAF has~\cite{johnston2000thermodynamics} logarithmic contributions to $\chi(T)$.

\begin{figure}[hbtp]
\begin{center}
\includegraphics[width=\columnwidth]{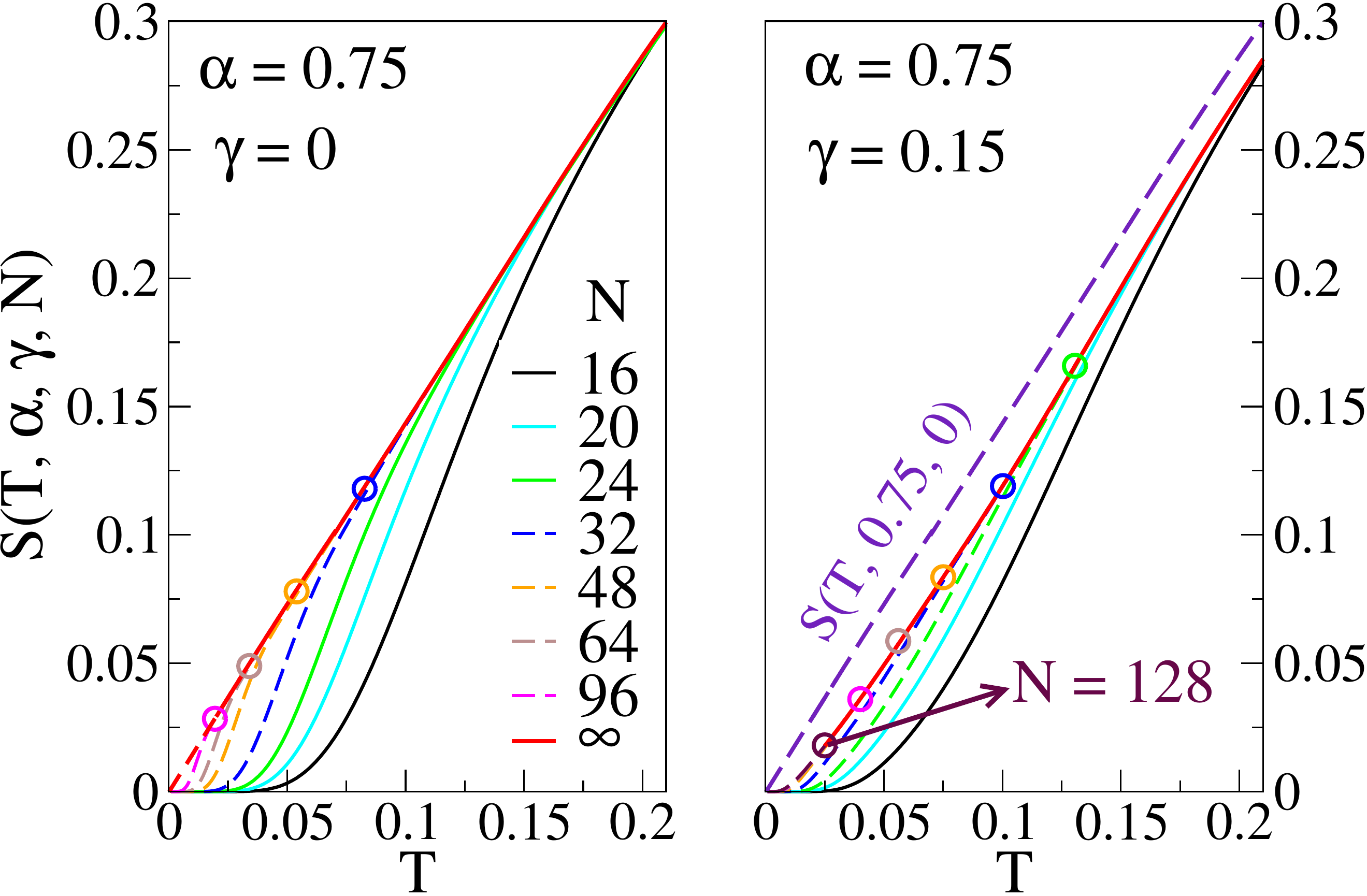}
\caption{Entropy per site, $S(T,\alpha,\gamma,N)$ of Eq.~(\ref{Eq3}) with $\delta = 0$, $N$ spins, $\alpha = 0.75$, for the $J_1-J_2$ model ($\gamma = 0$) and the extended ($\gamma = 0.15$) model. Solid lines are ED up to $N = 24$ for $\gamma = 0$ or $20$ for $\gamma = 0.15$ and converged DMRG for $T > T(N)$, shown as points; dashed lines for $T < T(N)$ are $S(T,\alpha,\gamma,N)$. The gap $ \Delta(\alpha, \gamma) $ reduces the $\gamma = 0.15$ entropy in the thermodynamic limit compared to $\gamma = 0$.}
\label{fig3}
\end{center}
\end{figure}

\begin{table}[hbtp]
\caption{\label{tab2}
	Size dependence of $T(\alpha,\gamma,N)$ of chains with $\alpha_{1}= 0.75$, $ \alpha_{2}=1.0 $ and $\gamma_1 = 0$, $\gamma_2 = 0.15$ in Eq.~(\ref{Eq3}) with $\delta = 0$.}
\begin{ruledtabular}
\begin{tabular}{c c c c c}
$N$ & $T(\alpha_{1}, \gamma_{1})$ & $ T(\alpha_{1}, \gamma_{2}) $ & $ T(\alpha_{2}, \gamma_{1}) $ & $ T(\alpha_{2}, \gamma_{2}) $ \\
\hline
32  & 0.091 & 0.101 & 0.096 & 0.125\\
48  & 0.062 & 0.075 & 0.054 & 0.078\\
64  & 0.042 & 0.062 & 0.037 & 0.056\\
96  & 0.025 & 0.040 & 0.028 & 0.039\\
\end{tabular}
\end{ruledtabular}
\end{table}
Figure~\ref{fig3} contrasts $S(T,\alpha,\gamma,N)$ at $\alpha = 0.75$ with $\gamma = 0$ and $0.15$ in the left and right panels. Solid and dashed lines are ED and DMRG, respectively; the points are $T(N)$ up to $N = 96$ for $\gamma = 0$ and $N = 128$ for $\gamma = 0.15$. The thermodynamic limit is obtained by interpolation between the points $T(N)$ and  ED of small systems. Linear extrapolation yields the $\gamma = 0$ entropy $S(T,\alpha,0)$ and finite $S^{\prime}(T,\alpha,0) = C(T,\alpha,0)/T$ at the origin; the exponentially small $\Delta(\alpha)$ presumably generates deviations from linearity at a decade or two lower $T$. The $\gamma = 0.15$ chain is clearly gapped, with $S^{\prime}(T) = 0$ at $T = 0$ and reduced entropy compared to $\gamma = 0$.

\begin{figure}[hbtp]
\begin{center}
\includegraphics[width=\columnwidth]{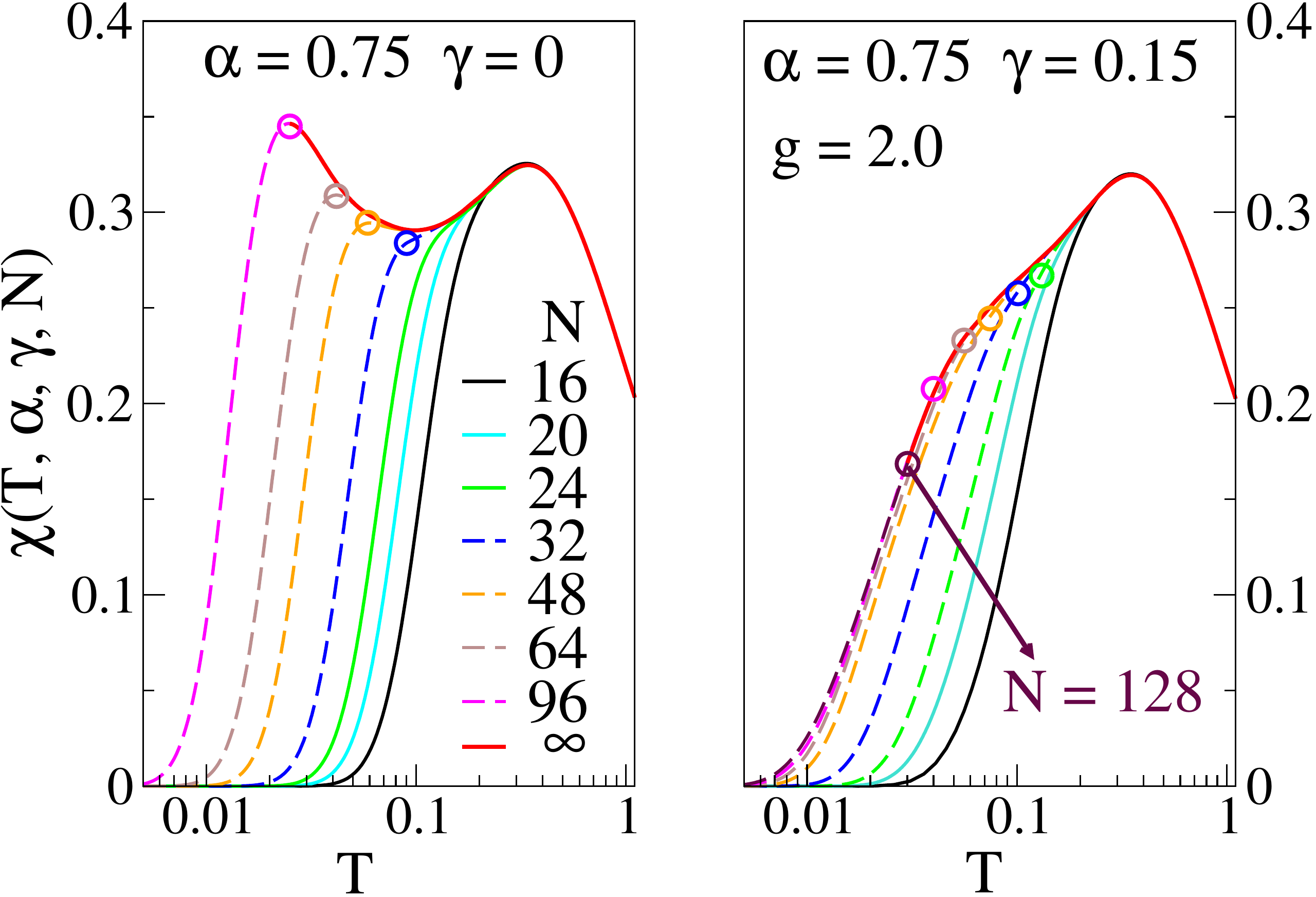}
\caption{Magnetic susceptibility, $ \chi(T,\alpha,\gamma,N) $, of Eq.~(\ref{Eq1}) with
$\delta = \gamma$,
$N$ spins, $g$ factor $2$, $\alpha = 0.75$ and $\gamma = 0$ ($J_1-J_2$ model) or $\gamma 
= 0.15$ (extended model). Solid lines are ED up to $N = 24$ for $\gamma = 0$ or $20$ for $\gamma = 0.15$ and converged DMRG for $T > T(N)$, shown as points; dashed lines for $T < T(N)$ are $\chi(T,\alpha,\gamma,N)$. The $T = 0$ susceptibilities are zero due to finite size gaps that converge to finite $\Delta(\alpha,\gamma)$ for $\gamma = 0.15$.}
\label{fig4}
\end{center}
\end{figure}

The size dependence of $\chi(T,\alpha,\gamma,N)$ is compared in Fig.~\ref{fig4} for $\alpha = 0.75$ chains with $\gamma = 0$ and $0.15$. The logarithmic scale emphasizes low $T$. The susceptibilities merge at slightly higher $T$ than the points $T(N)$ based on the entropy. The thermodynamic limit of $\chi(T,\alpha,0)$ in the left panel has a maximum at $T_m = 0.34$, a minimum at lower $T$ and it increases as $T\rightarrow0$, at least until $T\sim \Delta(\alpha)$. The gapped $\gamma = 0.15$ system in the right panel has increasing susceptibility up to a maximum. Convergence at low $T$ with system size indicates that $ \Delta(\alpha,\gamma) $ suppresses finite-size effects by $N \sim 100$. The $\gamma = 0$ and $0.15$ susceptibilities become equal for $T > 0.60$ because the total exchange does not depend on $\gamma$. Larger $ \Delta(\alpha,\gamma) $ converges at higher $T$.

The thermodynamics of $\beta$-TeVO$_{4}$ was previously based on the $J_1-J_2$ model and ED for $N < 20$ spins.~\cite{savina2011magnetic,pregelj2015,weickert2016magnetic} Finite size effects then appear in $\chi(T)$ at $T \sim 0.15$ or $T \sim 6$ K for $-J_1 \sim 40$ K. They become very large in Fig.~\ref{fig4} when $T$ is less than the singlet-triplet gap at system size $N$. The size dependence at $\gamma = 0.15$, a gapped chain, indicates convergence at $N = 128$ to $\chi(T)$ for $T > 0.03$, and is a good approximation down to $T = 0$.

\begin{figure}[hbtp]
\begin{center}
\includegraphics[width=\columnwidth]{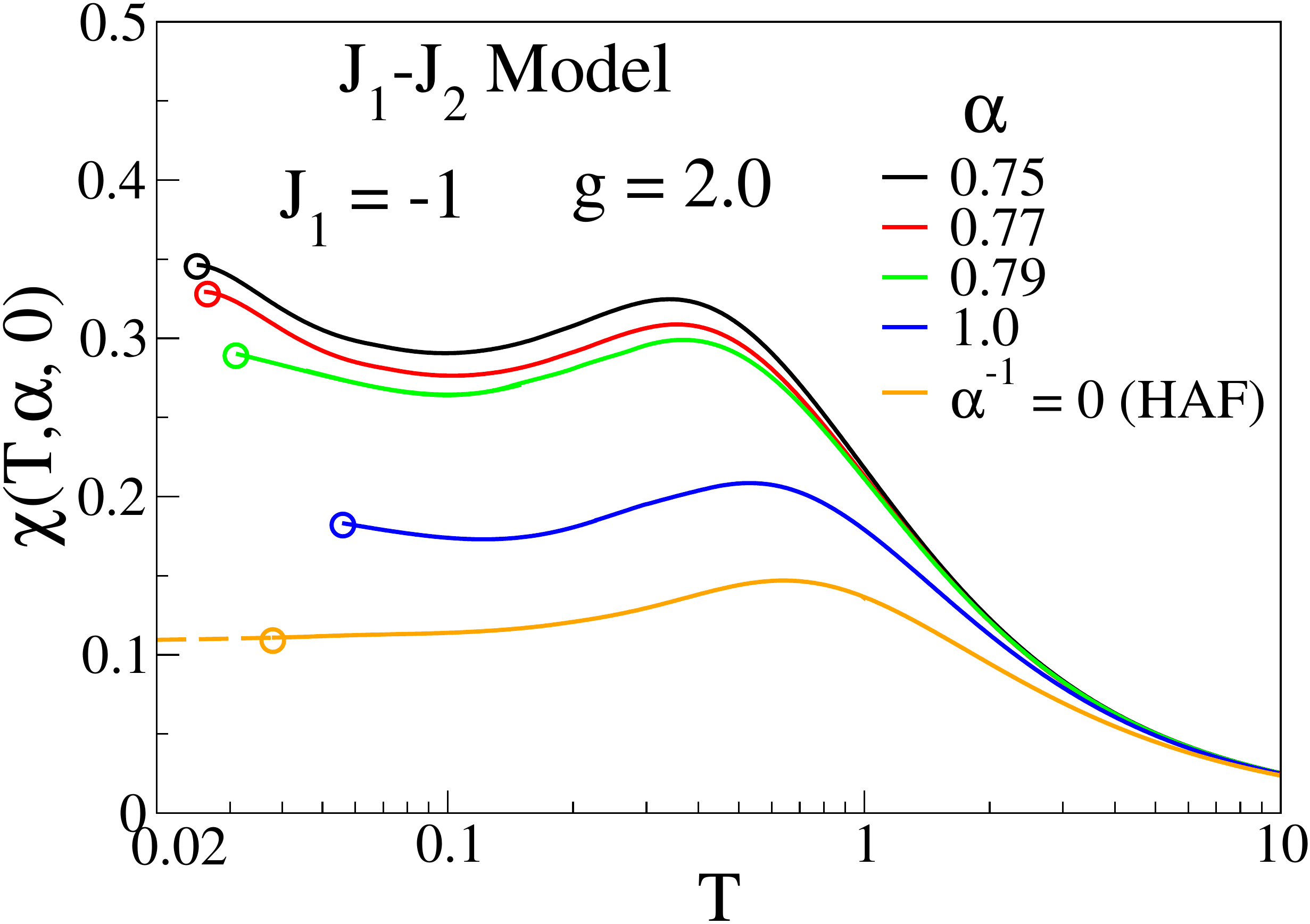}
\caption{Converged $ \chi(T,\alpha,0) $ of the $J_1-J_2$ model with $\delta = 0$ in Eq.~(\ref{Eq1}) for $T > T(\alpha,96)$, shown as points. The HAF susceptibility is extrapolated to lower $T$; it is exactly $1/\pi^{2} = 0.10132$ at $T = 0$.}
\label{fig5}
\end{center}
\end{figure}

Similar $\chi(T,\alpha,\gamma) $ or $ S(T,\alpha,\gamma) $ curves in the thermodynamic limit are obtained using ED for small $ N $, DMRG for the points $T(N)$ and extrapolation to $ T<T(N) $ for the largest $N$ studied. The $ \alpha $ dependence of $\chi(T,\alpha,0)$ is summarized in Fig.~\ref{fig5} for the ferromagnetic $J_1-J_2$ model with $\alpha > 0.65$. The smallest susceptibility is that of the HAF, whose extrapolation to lower $T$ agrees with the exact $\chi(T)$ for $T \geqslant 0.01$; logarithmic terms increase~\cite{johnston2000thermodynamics} the exact $\chi(0)=1/\pi^{2} $ by $6\%$ at $T = 0.005$. As expected on increasing antiferromagnetic exchange, $\chi(T,\alpha,0)$ decreases with increasing $ \alpha $. The $T$ dependence is monotonic at small $ \alpha $. It is almost a power law,~\cite{saha2019hybrid} $T^{-1.18}$, at $\alpha_{c} = 1/4$. There is a secondary maximum at $\alpha = 0.75$ that shifts to higher $T$ with increasing $ \alpha $ and becomes the only maximum for $\alpha > 1$.

\begin{figure}[hbtp]
\begin{center}
\includegraphics[width=\columnwidth]{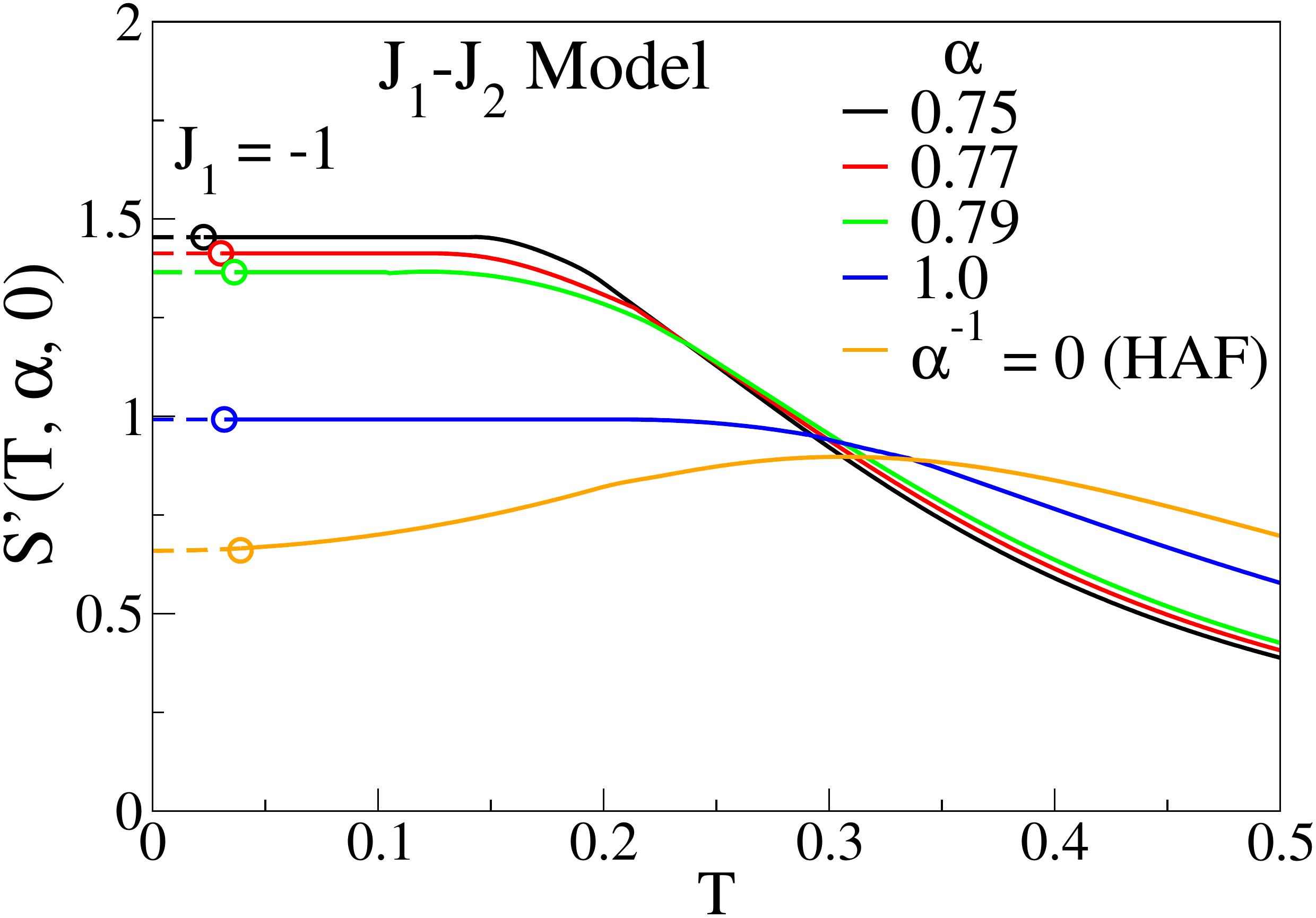}
\caption{Converged $S^{\prime}(T,\alpha,0) = C(T,\alpha,0)/T$ of the $J_1-J_2$ model with $\delta = 0$ in Eq.~(\ref{Eq1}) for $T > T(\alpha,96)$, shown as points.}
\label{fig6}
\end{center}
\end{figure}

The thermodynamic limit of $S^{\prime}(T,\alpha,0) = C(T,\alpha,0)/T$ is shown in Fig.~\ref{fig6} for the $J_1-J_2$ model at frustration $\alpha > 0.65$. We obtain $S^{\prime}(0) = 0.66$ for the HAF instead of the exact $2/3$. The density of low energy states increases with decreasing $ \alpha $. The maximum shifts to lower $T$ and merges into the $T = 0$ peak. The area under all curves is $ln 2$ in the high $T$ limit.

At fixed $ \vert J_{1}\vert=1 $, the ED/DMRG method yields the thermodynamics of $ H_{S}(\alpha, \gamma) $, Eq.~(\ref{Eq3}) with $\delta = 0$ to increasingly low $T \geqslant T(N)$ that depends on the system size $N$. The $\alpha > 0.65$ chains are unstable to sublattice dimerization. The equilibrium amplitude $ \delta(T) $ decreases from $ \delta(0) $ at $T = 0$ to $\delta(T_{SP}) = 0$ at the transition. The equilibrium $\delta(T)$ requires an explicit treatment of the lattice to balance the electronic instability.


\section{\label{sec4} Spin-Peierls transition}
The Peierls instability of a half-filled band of noninteracting fermions is driven by the filled valence band of bonding orbitals at $T = 0$. The conduction band of antibonding orbitals is empty. The curvature of the ground state diverges~\cite{longuet1959alternation} as $ E^{\prime \prime} \propto ln\delta $  and the band gap is $ 4\delta(0) $ in reduced ($t = 1$) units. Thermal population of antibonding orbitals and depopulation of bonding orbitals reduces $\delta(T)$. Dimerization is opposed by a harmonic lattice potential, $ \delta^{2}/2\epsilon_{d} $ per site, that corresponds to the optical phonon with wave vector $q = \pm \pi/2$. Two approximations are typically made in addition to linear electron-phonon coupling and a harmonic lattice.~\cite{su1980soliton} The adiabatic (Born-Oppenheimer) approximation neglects the nuclear kinetic energy; the mean-field approximation enforces equal amplitude $ \delta(T) $ at all bonds. The lattice force per site that opposes the electronic instability is $\delta/\epsilon_{d}$. The same approximations can be used for interacting fermions or correlated spin chains. 

The equilibrium sublattice dimerization for $ \alpha > 0.65$ in Eq.~(\ref{Eq3}) is
\begin{equation}
\frac{\delta(T)}{\epsilon_{d}} = - \left(\frac{\partial A(T,\delta)}{\partial \delta} \right)_{\delta(T)} \qquad T\leqslant T_{SP}
\label{Eq6}
\end{equation}
$A(T,\delta)$ is the free energy per site of the system under consideration. The thermodynamic limit of the $\delta$-derivative has long been known for noninteracting fermions via the grand canonical partition function, but not for correlated spin chains, not even for the HAF. Equation~(\ref{Eq6}) is the SP gap equation and a test of self consistency. Since $\delta(T_{SP}) = 0$, the observed $T_{SP}$ determines the stiffness $1/\epsilon_{d}$, or vice versa, and $\delta(T)$ down to $T=0$. One parameter, either $T_{SP}$ or the stiffness, governs the transition.

\begin{figure}[hbtp]
\begin{center}
\includegraphics[width=\columnwidth]{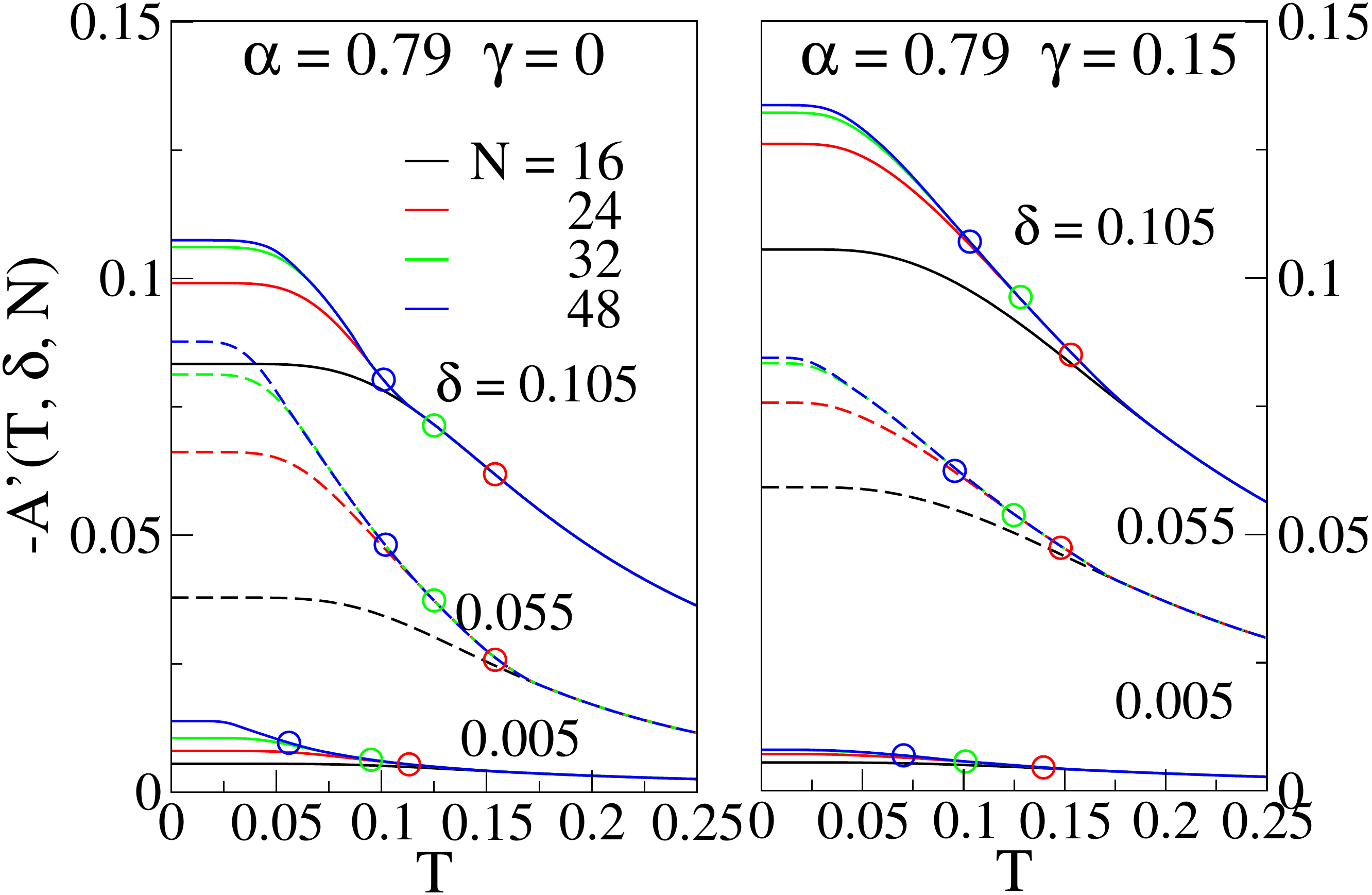}
\caption{The driving force $-A^{\prime}(T,\delta,N)$, Eq.~(\ref{Eq6}), for sublattice dimerization $\delta$ in the $J_1-J_2$ and extended $(\gamma = 0.15)$ models with $\alpha = 0.79$ and $N$ spins in Eq.~(\ref{Eq3}). Open circles are $T(N,\delta)$.}
\label{fig7}
\end{center}
\end{figure}

We consider $ A(T,\delta,N)=-TlnQ(T,\delta,N) $, where $ Q(T,\delta,N) $ is the canonical partition function at system size $N$, and solve Eq.~(\ref{Eq6}) for $\delta(T,N)$ for $T \leqslant T_{SP}$. Convergence of $ A^{\prime}(T,\delta,N)=\partial A(T,\delta,N)/\partial\delta $ to the thermodynamic limit holds for $T > T(N)$ as discussed in Sec.~\ref{sec3}. The size dependence of $A^{\prime}(T,\delta,N)$ is shown in Fig.~\ref{fig7} for $\alpha = 0.79$ and $\gamma = 0$ or $0.15$ in the left or right panels.

The driving force for sublattice dimerization, $-A^{\prime}(T,\delta,N)$, decreases with $T$ as expected. Its size dependence decreases with $ \delta $ since dimerization increases the gap $\Delta(\delta).$ At $\delta = 0.105$, the $A^{\prime}(T,\delta,N)$ in either panel have almost converged at $T = 0$ by $N = 32$. Close to $T_{SP}$, the $\delta = 0.005$ curves have converged to the thermodynamic limit for $T > T(32)$. The thermodynamic limit is reached at system size $N = 32$ at $T \sim T_{SP}$ due to high $T$ and at $T \sim 0$ due to large $\Delta(\delta(0))$.

\begin{figure}[hbtp]
\begin{center}
\includegraphics[width=\columnwidth]{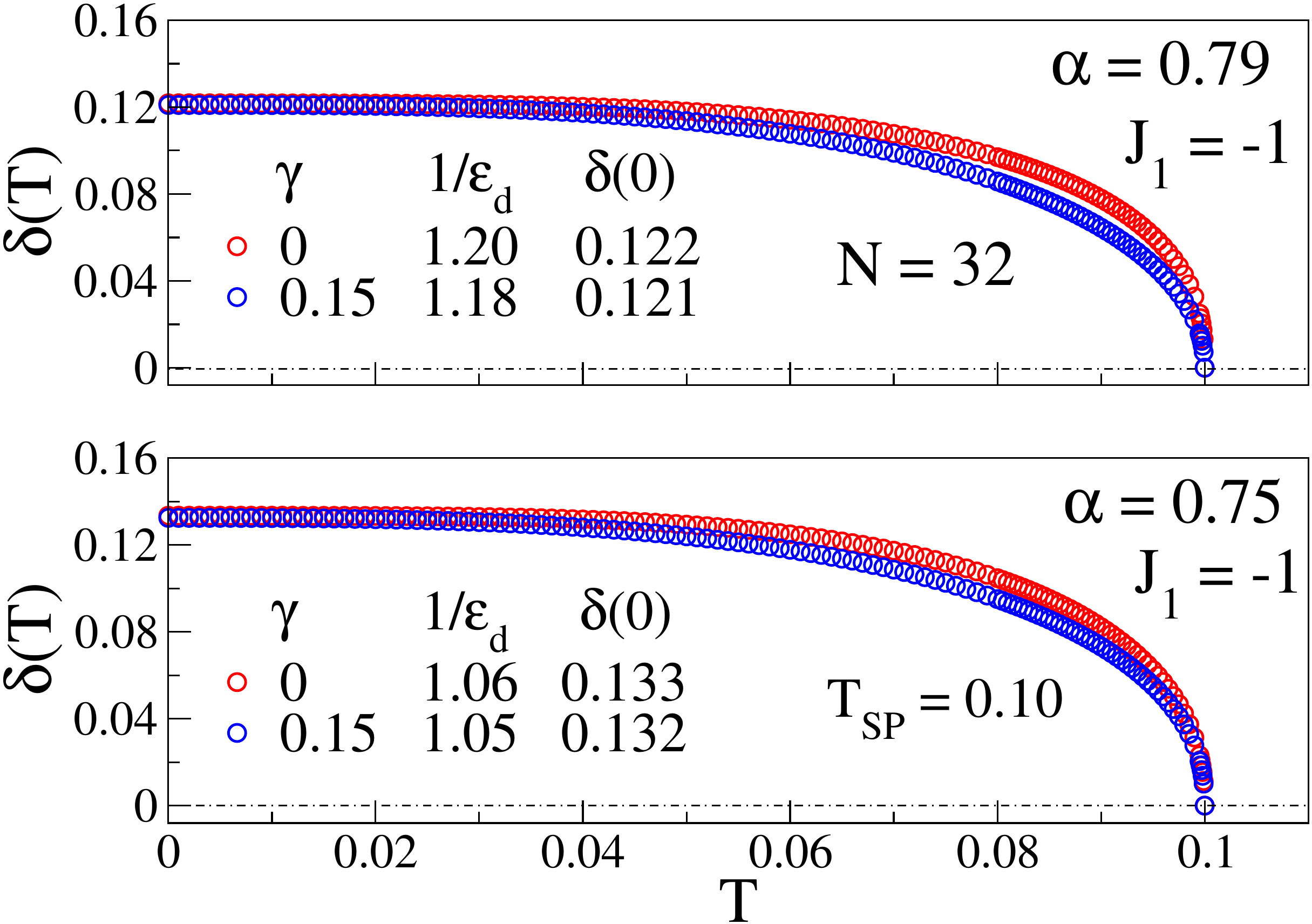}
\caption{Amplitude $\delta(T)$ of the converged sublattice dimerization, Eq.~(\ref{Eq7}), in spin chains with $T_{SP} = 0.10$ and $N = 32$ spins in Eq.~(\ref{Eq3}) with the indicated parameters.}
\label{fig8}
\end{center}
\end{figure}

Since $ \delta(T_{SP})=0 $, the curvature, $-A^{\prime \prime}(T_{SP},0)$, determines the stiffness parameter $1/\epsilon_{d}$ for a given $T_{SP}$. A typical value is $T_{SP} = 0.10$. Then $\delta(T)$ follows from Eq.~(\ref{Eq6}) for $T \leqslant T_{SP}$ and $\delta(0) = E^{\prime}(\delta(0))/A^{\prime \prime}(T_{SP},0)$. Almost identical $\delta(T,N)$ at $N = 24$ and $32$ indicates converged $\delta(T)$ at system size $N = 32$. Converged $\delta(T)$ are shown in Fig.~\ref{fig8} for $\alpha = 0.79$, $\gamma = 0$ and $0.15$ (upper panel) and $\alpha = 0.75$, $\gamma = 0$ and $0.15$ (lower panel). In order to evaluate $d\delta(T)/dT$, needed below, we fit $ \delta(T) $ to
\begin{equation}
\frac{\delta(T)}{\delta(0)}=\left[1-\left(\frac{T}{T_{SP}}\right)^{a}\right]^{b} \qquad   T\leqslant T_{SP}
\label{Eq7}
\end{equation}
The fitting parameters for $\alpha = 0.79$, $\gamma = 0$ are $a = 3.868$, $b = 0.354$; and for $\alpha = 0.75$, $\gamma = 0.15$, $a = 2.835$, $b = 0.408$. The $\delta(T)$ profile of the gapped $\gamma > 0$ chain is less abrupt near $T_{SP}$ than that of the gapless chain.

An SP transition generates a $C(T)$ anomaly and a discontinuous $\chi(T)$ slope at $T_{SP}$. We evaluate the anomaly for $T \leqslant T_{SP}$ as

\begin{eqnarray}
\frac{C(T)}{T} \equiv \frac{dS(T,\delta(T))}{dT} &=& \frac{\partial S(T,\delta(T))}{\partial T}\nonumber \\  &+& \frac{d\delta(T)}{dT}\frac{\partial S(T,\delta(T))}{\partial \delta}
\label{Eq8}
\end{eqnarray}
The first term is $C(T,\delta(T))/T$ at the equilibrium $\delta(T)$. The second term is discontinuous at $T_{SP}$ where $\delta(T) = 0$ for $T \geqslant T_{SP}$. As is well understood, the sharp discontinuity is due to the mean-field approximation that completely suppresses lattice fluctuations.

\begin{figure}[hbtp]
\begin{center}
\includegraphics[width=\columnwidth]{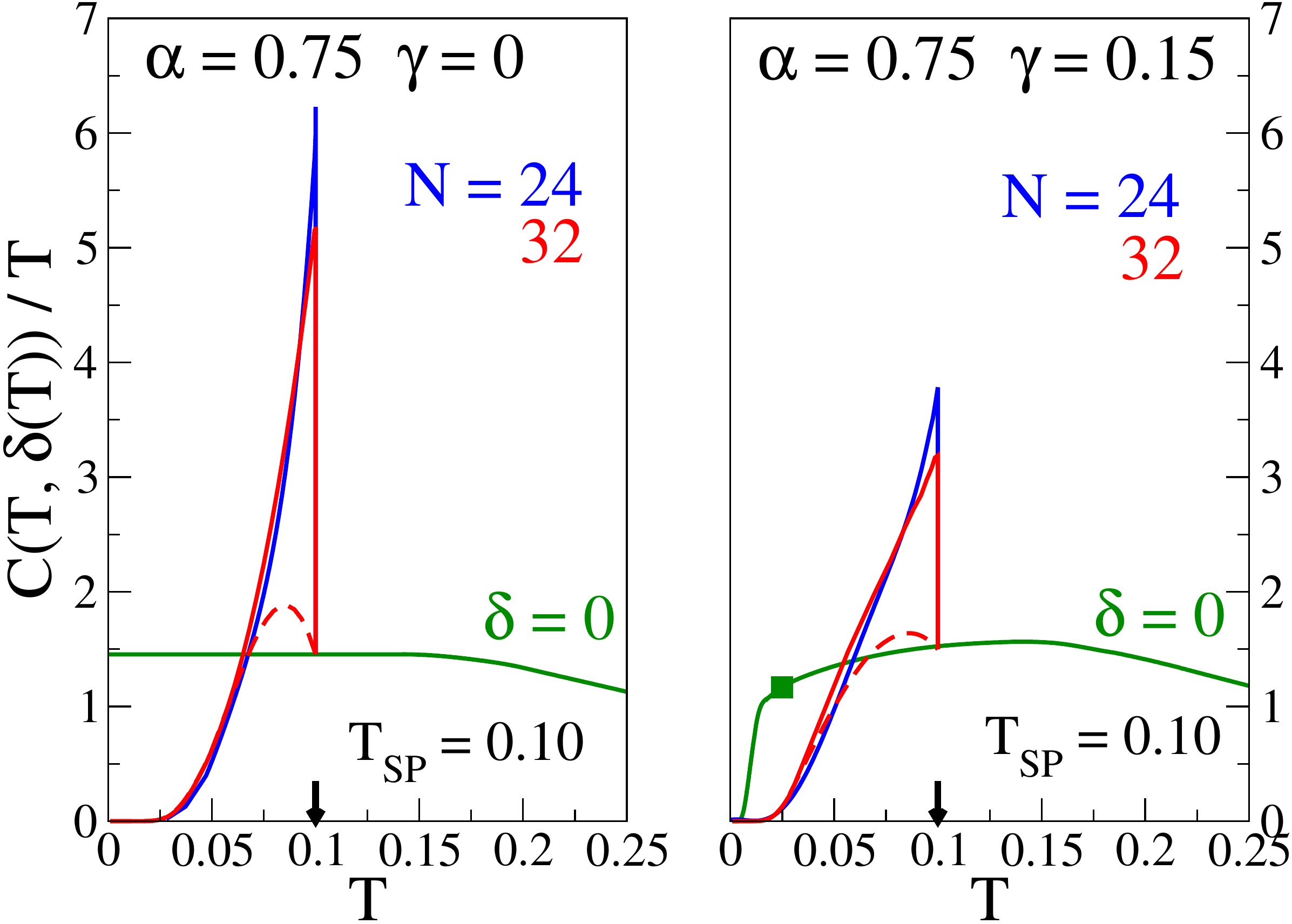}
\caption{ Calculated $C(T)/T$ for the SP transitions of the $J_1-J_2$ ($\gamma = 0$) and extended ($\gamma = 0.15$) models with $\alpha = 0.75$ and $T_{SP} = 0.10$. The continuous green ($\delta = 0$) curves are extended to $T = 0$ for $\gamma = 0$ and from $T(128)$ to $T = 0$ using Eq.~\ref{Eq9} for $\gamma = 0.15$. The solid blue and red lines are Eq.~\ref{Eq8} for $N = 24$ and $32$, respectively. The dashes red lines are $C(T,\delta(T))/T$, the first term of Eq.~\ref{Eq8} for $N = 32$.}
\label{fig9}
\end{center}
\end{figure}

Calculated $C(T)/T$ are shown in Fig.~\ref{fig9} for $T_{SP} = 0.10$, $\alpha = 0.75$ and $\gamma = 0$ (left panel), $\gamma = 0.15$ (right panel). The anomaly is based on system size $N = 32$, which has converged to the thermodynamic limit for these parameters. The continuous green ($\delta = 0$) lines are converged $C(T)/T$ for $T \geqslant T_{SP}$ or for systems without a transition. The green line $\delta = 0$ in the $\gamma = 0.15$ panel at $T < T_{SP}$ is based on extrapolating the $N = 128$ entropy according to
\begin{equation}
S(T)=c T^{-\eta} exp(-\Delta/T) \qquad  T \leqslant T(128)
\label{Eq9}
\end{equation}
Here $ \Delta $ is the singlet-triplet gap $\Delta(\alpha,\gamma)$, converged at $N = 128$. The exponent $ \eta $ and amplitude $c$ are obtained by equating $S(T)$ and $S^{\prime}(T)$ at $T = T(128)$ with the calculated magnitude and slope. The dashed red lines are $C(T,\delta(T))/T$, the first term of Eq.~(\ref{Eq8}), for the converged $N = 32$ systems.

As noted in Sec.~\ref{sec3}, the area under $S^{\prime}(T) = C(T)/T$ is $ln2$ in the high $T$ limit for all parameters. Since $C(T)/T$ is the same for $T \geqslant T_{SP}$, converged $S(T_{SP})$ immediately implies equal area up to $T_{SP}$ under the green $\delta = 0$ and red $\delta(T)$ curves in either panel of Fig.~\ref{fig9}. The curves cross at $T^{\prime} < T_{SP}$ and the difference between them up to $T^{\prime}$ is visibly smaller at $\gamma = 0.15$ than at $\gamma = 0$. The area from $T^{\prime}$ to $T_{SP}$ is then necessarily smaller for the gapped $\gamma = 0.15$ system with a conditional SP transition. We emphasize that the smaller $C(T)/T$ anomaly for $\gamma = 0.15$ is primarily due to the singlet-triplet gap $\Delta(\alpha, \gamma)$ and entropy $S(T_{SP},\alpha,\gamma)$. The lattice approximations leading to the calculated dimerization are secondary since $\delta(T)$ only changes the shape of the anomaly.

\begin{figure}[hbtp]
\begin{center}
\includegraphics[width=\columnwidth]{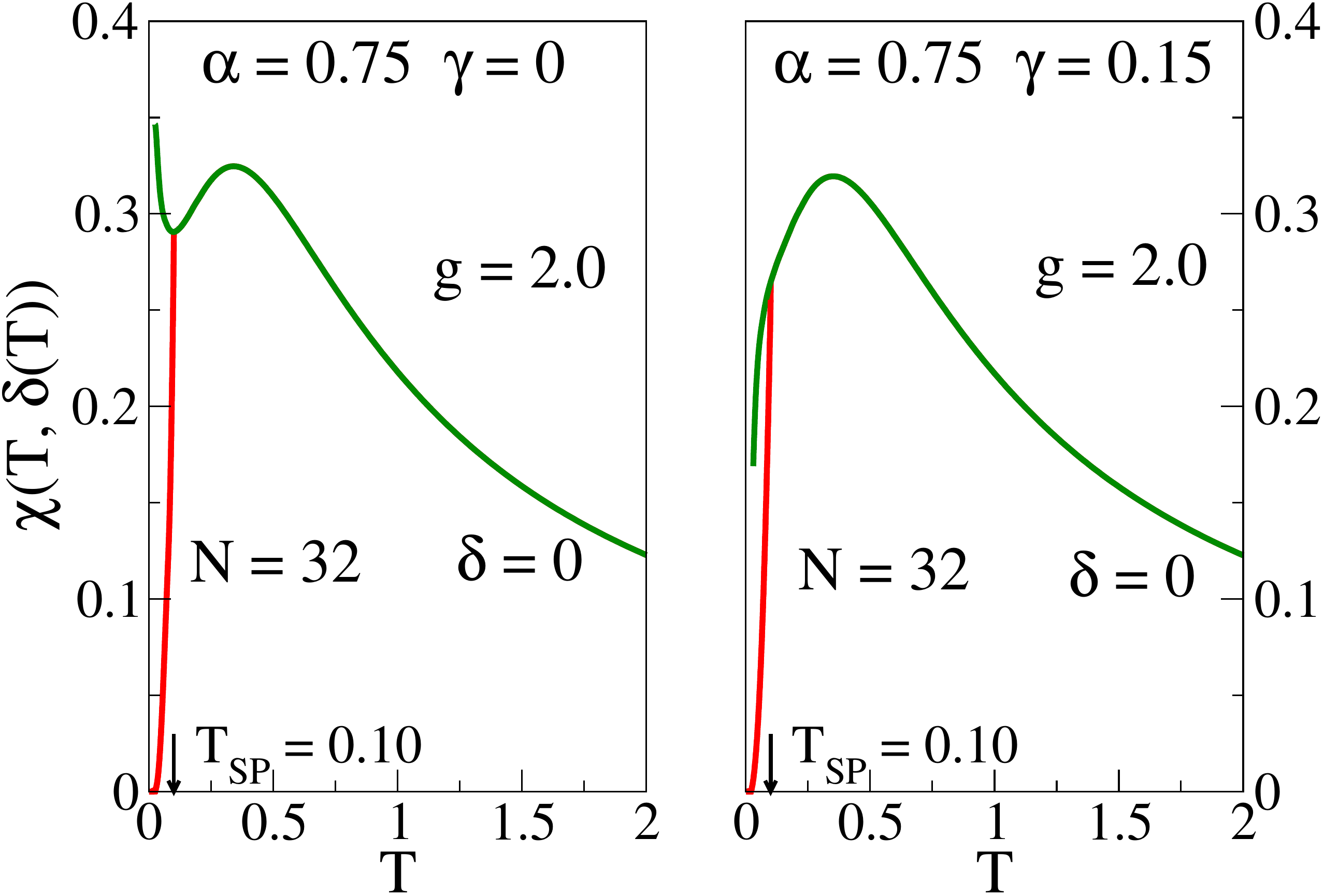}
\caption{ Calculated $ \chi(T,\delta(T)) $ of the $J_1-J_2$ ($\gamma = 0$) and extended ($\gamma = 0.15$) models with $\alpha = 0.75$ and $T_{SP} = 0.10$. The continuous green ($\delta = 0$) curves are converged $\chi(T,0)$ for $T \geqslant T(128)$. The red lines are $\chi(T,\delta(T))$ with the equilibrium $\delta(T)$ and are in the thermodynamic limit for $N = 32$.}
\label{fig10}
\end{center}
\end{figure}

Fig.~\ref{fig10} shows the calculated $ \chi(T,\delta(T)) $ for the models in Fig.~\ref{fig9}, with $T_{SP} = 0.10$, $\alpha = 0.75$ and $\gamma = 0$ (left panel), $\gamma = 0.15$ (right panel), and $\delta = 0$ for $T > T_{SP}$. The green ($\delta = 0$) curves are the converged susceptibilities of chains without an SP transition. The red curves are $\chi(T,\delta(T))$ for $T \leqslant T_{SP}$ with the equilibrium $\delta(T)$. The SP transition generates a discontinuous $ \chi(T) $ slope at $T_{SP}$. The large cusp in the $\gamma = 0$ panel is from a gapped chain with dimerized sublattices to a gapless chain. The cusp of the $ \gamma = 0.15 $ chain is reduced since sublattice dimerization merely increases the $\delta = 0$ gap. Approximations that suppress lattice fluctuations sharpen the $\chi(T)$ cusp.

We studied other SP transitions in chains with $\alpha > 0.65$ and $\gamma \geqslant 0$ in Eq.~(\ref{Eq3}) and found similar $C(T)$ and $\chi(T)$ results. The $\gamma > 0$ systems illustrate conditional instability to sublattice dimerization. For parameters leading to finite gaps $ \Delta(\alpha,\gamma) $, the SP transition combines a reduced but clear $C(T)$ anomaly with a modest $\chi(T)$ cusp at $T_{SP}$. We note that SP transitions have mostly been identified by $\chi(T)$ cusps, often without specific heat data.

\section{\label{sec5} $\beta$-T\lowercase{e}VO$_4$ Thermodynamics}
We have presented the thermodynamics of Eq.~(\ref{Eq3}) for $\alpha >0.65$ and its SP transition to sublattice dimerization. The extended $J_1-J_2$ model with isotropic exchanges has three parameters, $J_1$, $ \alpha $ and $ \gamma $. The fourth parameter, $T_{SP}$, specifies the stiffness $ 1/\epsilon_{d} $ of the harmonic lattice, or vice versa, and returns $\delta(T)$ for $T \leqslant T_{SP}$. The model, linear spin-phonon coupling and approximations for the lattice are well defined. The transition and thermodynamics are fully specified by four parameters.

The quasi-1D material $\beta$-TeVO$_{4}$ has zigzag chains along the $c$ axis. Smaller exchanges of either sign have been computed~\cite{saul2014density} between spins in adjacent chains, and interchain exchange has been invoked~\cite{pregelj2015,weickert2016magnetic,savina2015features} for the phase transitions at $2.3$ K, $3.3$ K and $4.6$ K. Aside from an early report, the magnetic susceptibility and spin specific heat at higher $T$ have been analyzed in terms of the $J_1-J_2$ model, $\delta = 0$ in Eq.~(\ref{Eq1}).

We list in Table~\ref{tab3} three published $ J_1 $ and $ \alpha $ that yield good $ \chi(T) $ and $C(T)$ fits for $T > 8$ K in the paramagnetic phase and include two extended models with $J_1$, $ \alpha $ and $\gamma = 0.15$. Also listed are the calculated entropy $S(T)$ at $T = 5.5$ K. The energy per site of the ferromagnetic phase with fully aligned spins in a uniform applied magnetic field B is
\begin{equation}
E_{F}(\alpha, \gamma, B) = -\frac{(1-\alpha)}{4}-\frac{hB}{2}
\label{Eq10}
\end{equation}
with $h = g\mu_{B}/\vert J_{1}\vert $, $g = 2$ and Bohr magneton $\mu_{B}$. $B_{sat}$ in Table~\ref{tab3} is the field at which the absolute ground state at $0$ K becomes ferromagnetic. Other measurements in the paramagnetic phase provide additional applications of $H(\alpha,\gamma)$. 

\begin{figure}[hbtp]
\begin{center}
\includegraphics[width=\columnwidth]{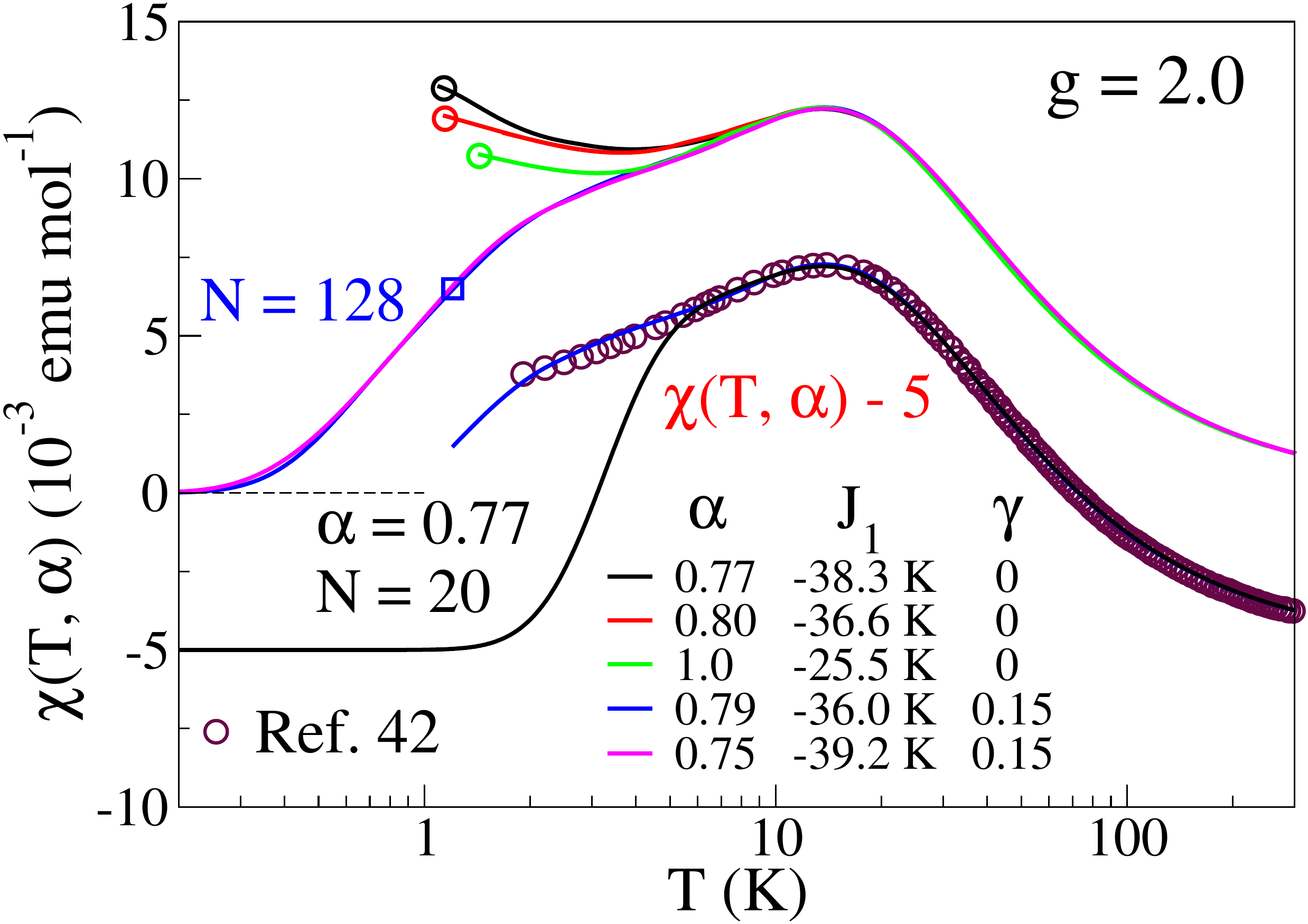}
\caption{ Molar $\chi(T)$ of $\beta$-TeVO$_4$. The data down to $T = 1.9$ K is from ~\cite{savina2011magnetic}. Except for ED at $N = 20$, the calculations are converged $\chi(T)$ with the indicated parameters. The $\chi(T) - 5$ curves compare experiment with ED at $N = 20$ and a converged $\gamma = 0.15$ model. The calculated $\chi(T)$ agree with each other for $T > 8$ K but disagree at lower T. Open circles are $T(96)$ for $J_1-J_2$ models with $\gamma = 0$; the square is $T(128)$ for either extended model.}
\label{fig11}
\end{center}
\end{figure}

\begin{table}[hbtp]
\caption{\label{tab3}
Parameters $ J_1 $ and $ \alpha $ in Ref. ~\cite{savina2011magnetic}, ~\cite{pregelj2015} and ~\cite{weickert2016magnetic} for $ \chi(T) $ fits of $\beta$-TeVO$_4$ in the paramagnetic phase. $B_{sat}$ and $S(5.5)$ are the calculated saturation magnetic field and the molar entropy at 5.5 K. In parentheses, the scaled $J_1$ discussed in the text and resulting $B_{sat}$ and $S(5.5)$. The extended $J_1-J_2$ models in the last two rows have $\gamma = 0.15$.}
\begin{threeparttable}
\begin{ruledtabular}
\begin{tabular}{c c c c}
$-J_1 $ (K) & $ \alpha = J_2/\vert J_1 \vert $ & $ B_{sat} $ (T) & $ S(5.5) $ (J/Kmol) \\
\colrule
38.3\tnote{a}        & 0.77 & 22.3        & 1.67\\
38.3\tnote{b} (36.6) & 0.80 & 24.5(23.4)  & 1.55(1.62)\\
26.6\tnote{c} (25.5) & 1.0  & 24.8(23.7)  & 1.56(1.58)\\
36.0\tnote{d}        & 0.79 & 23.4        & 1.43\\
39.2\tnote{d}        & 0.75 & 23.3 & 1.45 \\
\end{tabular}
\end{ruledtabular}
\begin{flushleft}
$^{a}$Reference~\cite{savina2011magnetic}\\
$^{b}$Reference~\cite{pregelj2015}\\
$^{c}$Reference~\cite{weickert2016magnetic}\\
$^{d}$ $ \gamma = 0.15 $
\end{flushleft}
\end{threeparttable}
\end{table}
The reported $\chi(T)$ are not quite identical but differ by a few percent. We took as reference the $ \chi(T) $ data of Savina et al.~\cite{savina2011magnetic} with the applied field along the $a$ or $c$ axis and $g = 2.0$; the range from $1.9$ to $6$ K is from Fig. 7 and higher $T$ is from Fig. 4. The measured $\chi(T)$ in Fig.~\ref{fig11} is shifted down by $ 5 \times 10^{-3} $ and compared to the calculated $\gamma = 0.15$ curves, also shifted down, which are almost identical. We include the exact $\chi(T,N)$ for $N = 20$, larger than $N = 18$ in Ref.~\cite{savina2011magnetic}, that deviates from experiment at low $T$ due to finite size gaps.

Since the present context calls for model-to-model comparisons and calculations, we rescaled the other data by changing $-J_1$ from $38.3$ K to $36.6$ K in Ref.~\cite{pregelj2015} and from $26.6$ K to $25.5$ K in Ref.~\cite{weickert2016magnetic}. The scaled $J_1$ in Table ~\ref{tab3} have slightly reduced $B_{sat}$ and increased $S(5.5)$. The calculated $\chi(T)$ in Fig.~\ref{fig11} are then in good agreement with each other in the paramagnetic phase. Convergence at $T > 8 K$ demonstrates the insensitivity of $\chi(T)$ to matched parameters $J_1$, $ \alpha $ and $ \gamma $ in Table ~\ref{tab3}.

The opposite holds at low $T$, which brings out clear differences among the calculated $ \chi(T) $. The solid lines in Fig.~\ref{fig11} are in the thermodynamic limit for $T > T(N)$, the indicated points for $N = 96$ for the $J_1-J_2$ models and $N = 128$ for the extended models with $\gamma = 0.15$. Fig.~\ref{fig4} shows the convergence of $\chi(T,N)$ to the thermodynamic limit with increasing $N$. The gap $\Delta(\alpha,\gamma) \sim 0.05 \vert J_1\vert$ at $\gamma = 0.15$ has almost converged as is evident from the $N = 96$ and $128$ curves. Much lower $T$ is needed to probe the far smaller $\Delta(\alpha)$ of $J_1-J_2$ models in the IC phase. We conclude that the $J_1-J_2$ model with $\delta = 0$ in Eq.~(\ref{Eq1}) does not account for the $\beta$-TeVO$_{4}$ susceptibility below $8$ K.

\begin{figure}[hbtp]
\begin{center}
\includegraphics[width=\columnwidth]{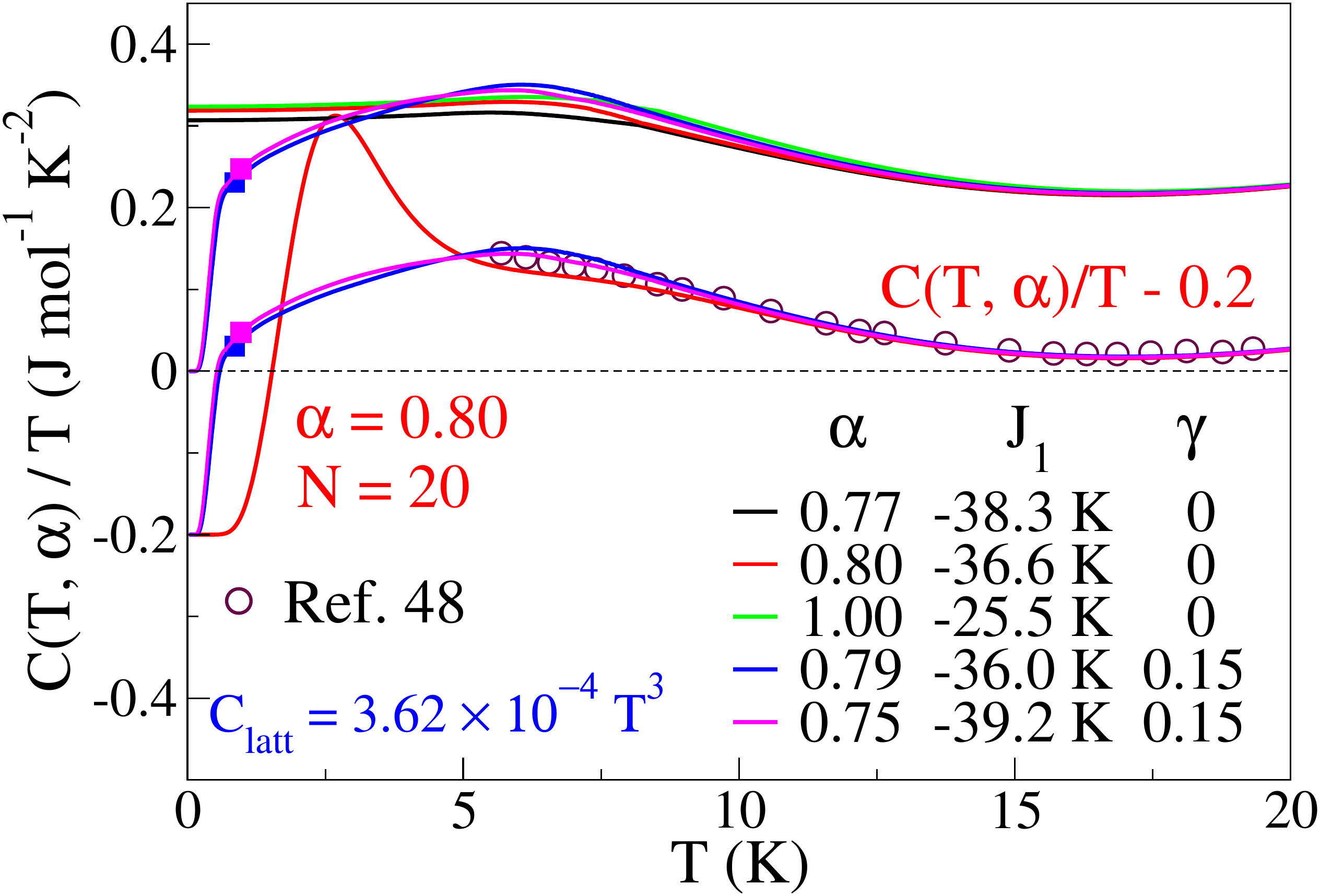}
	\caption{Molar $C(T)/T$ of $\beta$-TeVO$_{4}$ after subtracting a (Debye) lattice contribution of $C_{latt}$. The data is from Ref. ~\cite{savina2015heat}. The calculations are converged $C(T)/T$ except for ED at $N = 20$. The calculated curves agree with each other for $T > 7$ K but not at lower $T$. The $J_1-J_2$ ($\gamma = 0$) curves are finite at $T \sim 0$. The $\gamma = 0.15$ lines for $T < T(128)$ are extrapolations according to Eq.~(\ref{Eq9}).}
\label{fig12}
\end{center}
\end{figure}

The specific heat is dominated by lattice phonons for $T > 15$ K. The Debye $T^{3}$ law for acoustic phonons and mass $M$ per unit cell is a low-$T$ approximation for $\beta$-TeVO$_4$ with four formula units per unit cell. Moreover, $C(T)$ is more sensitive to model parameters than $\chi(T)$. The $C(T)$ data for $T > 8$ K in Fig.~\ref{fig12} is from Fig. 1 of Ref.~\cite{savina2015heat}, shifted down by $0.2$, and compared to the calculated $\gamma = 0.15$ specific heats, also shifted, and $C_{latt}(T) = 3.6 \times 10^{-4}$ $T^{3}$. Finite size gaps are evident in the exact $N = 20$ curve at low $T$. Entropy conservation ensures that $C(T,N)/T$ converges to the thermodynamic limit from above since reduced $S(T,N)$ at low $T$ must be offset by increased $S(T,N)$ before converging to $S(T)$. Previous modeling of $C(T)$ and $\chi(T)$ in the paramagnetic phase with $N < 20$ is inadequate at low $T$.

The calculated molar $C(T)/T$ in Fig.~\ref{fig12} are for the same parameters as in Fig.~\ref{fig11}. The $\gamma = 0.15$ lines are converged for $T\geqslant T(128)$ and
are based on Eq.~(\ref{Eq9}) at lower $T$. As discussed in Sec.~\ref{sec4}, the $J_1-J_2$ curves remain finite at least down to $T \sim \Delta(\alpha)$, the exponentially small gap in the IC phase, while the gapped $ \gamma = 0.15 $ chains have $S^{\prime}(0) = 0$. There is fair convergence for $T > 7$ K and excellent convergence at $T > 10$ K, fully comparable to the $\chi(T)$ convergence. The thermodynamic limit of $C(T)$ is also quite different at low $T$ for the model parameters in Table~\ref{tab3}.

\begin{figure}[hbtp]
\begin{center}
\includegraphics[width=\columnwidth]{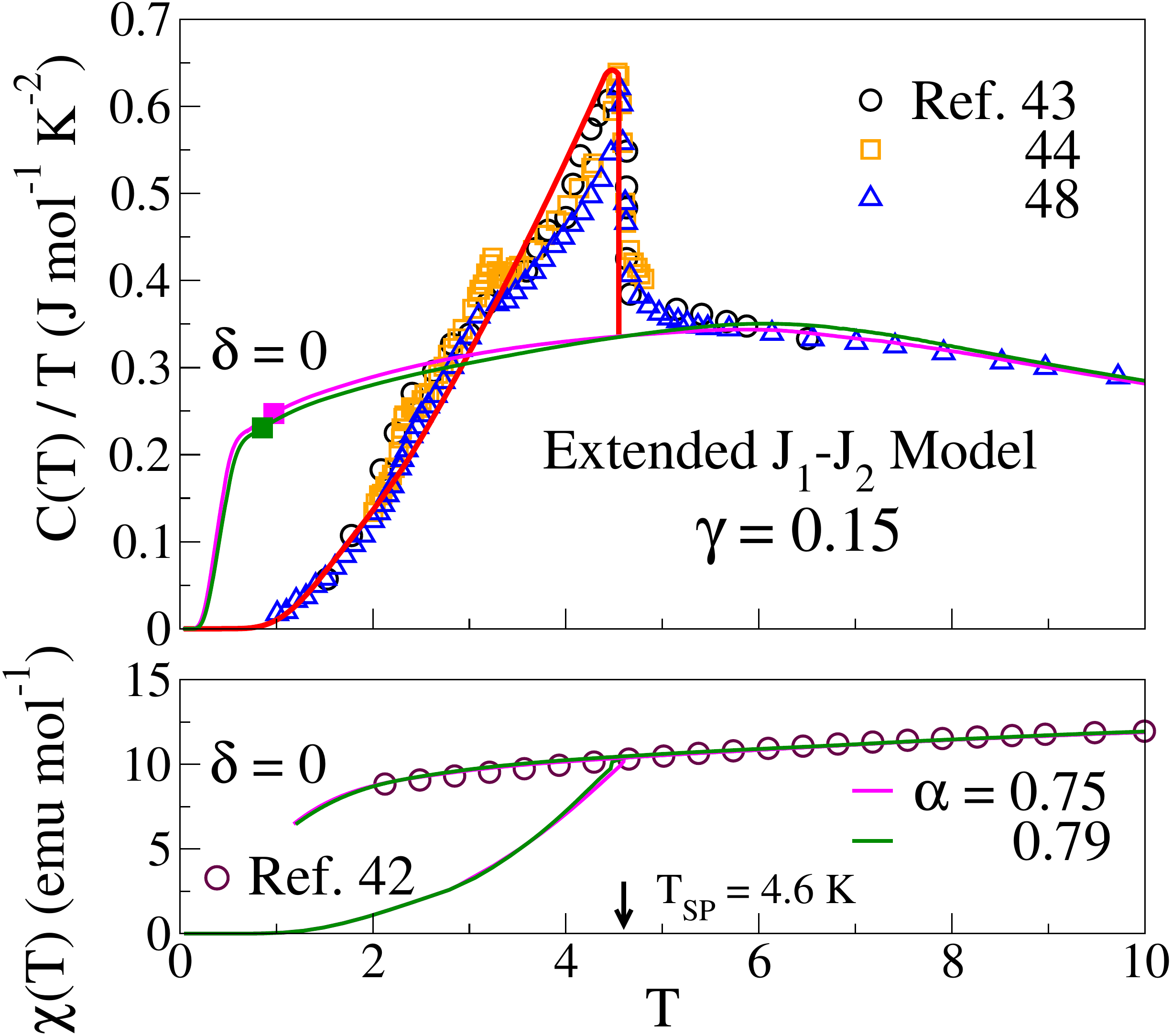}
\caption{Upper panel: Measured $C(T)/T$ from Refs.~\cite{pregelj2015} and ~\cite{weickert2016magnetic} for $T < 6$ K and from Ref.~\cite{savina2015heat} for $T < 10$ K. The $\delta = 0$ curves are the calculated $\gamma = 0.15$ curves in Fig.~\ref{fig12}. The red line is the calculated conditional SP transition for either system with $\gamma = 0.15$ using Eq.~(\ref{Eq6}). Lower panel: Measured $\chi(T)$ from Ref.~\cite{savina2011magnetic}. The $\delta = 0$ curves are the calculated $\gamma = 0.15$ curves in Fig.~\ref{fig11}. The $\chi(T,\delta(T))$ curves with equilibrium $\delta(T)$ are for extended models with $\gamma = 0.15$.} 
\label{fig13}
\end{center}
\end{figure}

The $\beta$-TeVO$_4$ thermodynamics discussed so far show that all five models in Table~\ref{tab3} account for $\chi(T)$ and $C(T)$ data at $T > 8$ K. The extended models with $\gamma = 0.15$ are clearly superior in Fig.~\ref{fig11} for $\chi(T)$ at $T < 8$ K. Three groups have reported $C(T)$ measurements~\cite{pregelj2015,weickert2016magnetic,savina2015heat} that include the $4.6$ K transition. The data is plotted as $C(T)/T$ in the upper panel of Fig.~\ref{fig13}. The anomaly is broadened on the high-T side, as expected qualitatively on the basis of lattice fluctuations.

The integral of the $C(T)/T$ data up to $T = 5.5$ K is the measured entropy $S(5.5)$. The Debye contribution is negligible here. We evaluated the areas as $1.35$, $1.43$ and $1.47$ J/Kmol in Ref.~\cite{savina2015heat}, ~\cite{weickert2016magnetic} and ~\cite{pregelj2015}, respectively. The calculated $S(5.5)$ for $\gamma = 0$ in Table~\ref{tab3} are about $10\%$ greater than measured. The adjusted $J_1$ mentioned in connection with $\chi(T)$ increase $S(5.5)$ by $0.07$ and $0.02$ J/Kmol for Refs.~\cite{pregelj2015} and ~\cite{weickert2016magnetic}. Since entropy is a state function, $S(5.5)$ is independent of the path from $T = 0$ to $5.$5 K. The $5.5$ K entropy motivated our choice of the extended model with $\gamma = 0.15$.

The solid lines labeled $\delta = 0$ in the upper panel of Fig.~\ref{fig13} are the calculated $C(T)/T$ of the $\gamma = 0.15$ models, also shown in Fig.~\ref{fig12}. The same parameters generate the red curve when the anomaly is a conditional SP transition at $4.6$ K. The absolute specific heat is fit semi-quantitatively on taking $T_{SP} = 4.6$ K. The larger anomaly of gapless $J_1-J_2$ models with unconditional transitions is shown in Fig.~\ref{fig9}.

The molar spin susceptibility $\chi(T)$ at $T < 10$ K is shown in the lower panel of Fig.~\ref{fig13}. The data and the curve labeled $\delta = 0$ are the same as in Fig.~\ref{fig11}. The extended models with $\gamma = 0.15$ in Table~\ref{tab3} work well for $\chi(T)$ over the entire range to $300$ K. The calculated $\chi(T,\delta(T))$ for same parameters and $T_{SP} = 4.6$ K have a cusp at the transition. The slope change of $\chi(T)$ at $T_{SP}$ is much larger than the tiny signature in Fig. 7 of Ref.~\cite{savina2011magnetic}.

A remarkable feature of the $4.6$ K transition is a clear $C(T)/T$ anomaly with hardly any $\chi(T)$ signature. We have of course explored other parameters. The fundamental issue is that the $C(T)/T$ data require a larger gap at low $T$ than in $\delta = 0$ models that yield the thermodynamics at higher $T$. However, an increased gap necessarily reduces $\chi(T)$. The reconciliation of a small $\chi(T)$ gap and large $C(T)$ gap is a serious problem for quantitative modeling of $\beta$-TeVO$_{4}$.

We emphasized in the Introduction that 1D models with isotropic exchange such as Eq.~(\ref{Eq1}) are first approximations to the magnetic properties of quasi-1D materials. The saturation field $B_{sat}$ is another example for $\beta$-TeVO$_4$. The $g$ factor~\cite{savina2011magnetic,pregelj2015} is close to $2.0$ when $B$ is along the $a$ or $c$ axis. The calculated $B_{sat}$ in Table~\ref{tab3} is necessarily the applied field for models with only isotropic exchange. Two measurements ~\cite{osti_1345914,pregelj2020thermal} at $T \sim 50$ mK returned $B_{sat} \sim 21.7$ T. The difference can plausibly be attributed to an internal field that augments $B_{app}$ even if the origin and magnitude of the field are open at present. There are other reasons for supposing an internal field.

$\beta$-TeVO$_{4}$ modeling beyond the $J_1-J_2$ model has already been considered qualitatively. The low symmetry of zigzag chains does not rule out either antisymmetric or anisotropic corrections to isotropic exchange. There are interchain exchanges $J^{\prime}$ of either sign between nearby spins in different chains. But specific extensions of the $J_1-J_2$ model have yet to be identified and are likely to be challenging for quantitative analysis.

\section{\label{sec6} Discussion}
The spin-Peierls instability of the ferromagnetic $J_1-J_2$ model in the singlet sector
of Eq.~(\ref{Eq1}) with $J_1 < 0$ and $\delta = 0$ differs in several ways from SP systems with $J_1 > 0$. The instability for $\alpha = J_2/\vert J_1\vert > 0.65$ is \textit{sublattice} dimerization to singlet ground states with four spins per unit cell. Smaller $ \alpha $ leads to \textit{chain} dimerization, as usual for $J_1 > 0$, with two spins per unit cell down to $\alpha_{c} = 1/4$. Our discussion of the SP transitions to dimerized sublattices of the $J_1-J_2$ model is otherwise standard: linear coupling, harmonic lattice, adiabatic and mean-field approximations for the lattice. The SP instability of the ferromagnetic $J_1-J_2$ model is unconditional in the singlet sector.

The extended model with $\delta = 0 $ in Eq.~(\ref{Eq3}) has lower symmetry, a nondegenerate singlet ground state for the parameters of interest, and a finite singlet-triplet gap $ \Delta(\alpha,\gamma) $. The SP instability is conditional and the transition in a suitably soft system with $\alpha > 0.65$ is to sublattice dimerization. A conditional SP transition reduces the magnitude of the $C(T)/T$ anomaly and the $\chi(T)$ cusp at $T_{SP}$, as shown in Sec.~\ref{sec4}.

ED/DMRG is a recent approach~\cite{saha2019hybrid} to the low T thermodynamics of 1D spin chains. It has been validated against some exact and numerical results,~\cite{saha2021} but its acceptance as an accurate technique will depend on additional comparisons. The SP transition of CuGeO$_{3}$ has been modeled by TMRG~\cite{Raupach_1999} and ED/DMRG~\cite{saha_2020} using identical approximations for the lattice and the same parameters $J_1=160$ K, frustration $\alpha =0.35$ and $T_{SP} = 14.4 $ K. With the TMRG results first, the stiffness $K$ (or $1/\epsilon_d$) $= 11J_1 (11.1J_1)$, $\delta(0)=0.026$ ($0.0248$) and gap $\Delta = 40$ K ($38$ K) at $T = 0 $ K. The same $\chi(T)$ data was fit quantitively for $T > T_{SP}$ in Fig. 4 of~\cite{Raupach_1999} and over the entire range in Fig. 5 of~\cite{saha_2020}, as required by Eq. ~(\ref{Eq6}) for self-consistency. A quantitative TMRG fit of $\chi(T)$ below $T_{SP}$ required~\cite{Raupach_1999} a softer $K = 10.2J$ leading to higher $T_{SP}=15.2$ K. Self-consistency is satisfied to better than $8\%$. The mean field fit of $\chi(T)$ below $T_{SP}$ in the organic SP system with $\alpha = 0$ leads to~\cite{jacobs1976spin} $T_{SP} = 9$ K, $25\%$ less than the observed $12$ K. 

The ED/DMRG method makes accessible the low $T$ thermodynamics of correlated spin chains such as the $J_1-J_2$ and extended models. System sizes of $N \sim 100$ or larger are needed to suppress finite-size effects. Since spin correlations decrease at high $T$, small $N$ becomes sufficient. ED for $J_1-J_2$ models with $N < 20$ and several combinations of $J_1$ and $\alpha$ account for the magnetic susceptibility $ \chi(T) $ and spin specific heat $C(T)$ of $\beta$-TeVO$_{4}$ in the paramagnetic phase, but fail below $10$ K because the data indicate a gapped system. The extended models with $\gamma = 0.15$ and the $J_1$, $ \alpha $ combinations in Table~\ref{tab3} are then required for reasons given in Sec.~\ref{sec5}.

We have not been able to model the $C(T)/T$ and $\chi(T)$ data in Fig.~\ref{fig13}. The $\delta = 0$ curves in the lower panel account for $\chi(T)$ down to 2 K without an SP transition at 4.6 K. The same $J_1$, $ \alpha $,  $ \gamma $ and $T_{SP} = 4.6$ K generate the red $\gamma = 0.15$ curves in the upper panel. But the large gap due to sublattice dimerization is incompatible with $\chi(T)$.

The $J_1-J_2$ model has previously been used to discuss the quantum phases of $\beta$-TeVO$_{4}$ at low $T$ and variable applied field from $B = 0$ to $B > B_{sat}$. The analysis is consistent but qualitative. The quantum phases reflect the symmetry of the $J_1-J_2$ model, which is far higher than the symmetry of the material. Both $\chi(T)$ and $C(T)$ point to gapped models with much larger gaps than $ \Delta(\alpha) $, although the gap is not necessarily due to $ \gamma $. Comparison with experiment returns accurate parameters of 1D models with isotropic exchange that are the starting point for detailed characterization of the magnetic properties of complex quasi-1D materials.

\begin{acknowledgments}
We thank the reviewer for bringing Ref.~\cite{Raupach_1999} to our attention. M.R. thanks S. N. Bose National Centre for Basic Sciences for fellowship. S.K.S. thanks DST-INSPIRE for financial support. M.K. thanks SERB for financial support through Grant Sanction No. CRG/2020/000754. 
\end{acknowledgments}


%
\end{document}